\documentclass[12pt, a4paper]{elsarticle}
\usepackage[letterpaper, portrait, margin=1in]{geometry}
\usepackage[hidelinks]{hyperref}

\usepackage{amssymb}

\usepackage{float}
\restylefloat{table}

\usepackage{physics} 
\usepackage{subcaption} 
\usepackage{ragged2e} 
\usepackage{mathtools} 
\usepackage{csquotes} 
\usepackage{xcolor}
\MakeOuterQuote{"}

\newcommand{\edit}[1]{{\color{black}#1}}
\def \ADP {\text{ADP}}
\def \kadh {k_\text{adh}}
\def \kcoh{k_\text{coh}}
\newcommand{\p}{\mathrm{p}}
\newcommand{\pt}{\tilde{\mathrm{p}}}
\def \Pmax {P_\text{max}}
\def \Pdiam {P_\text{diam}}

\def \n {\vu{n}}
\def \shear {\dot{\gamma}}
\def \u {\va{u}}
\def \v {\va{v}}
\def \x {\va{x}}

\journal{SoftwareX}

\begin{document}

\begin{frontmatter}



\title{\edit{clotFoam: An Open-Source Framework to Simulate Blood Clot Formation Under Arterial Flow}}


\author[label1]{David Montgomery}
\author[label2]{Federico Municchi}
\author[label3,label4]{Karin Leiderman}
\address[label1]{Department of Applied Mathematics and Statistics, Colorado School of Mines, 1500 Illinois St, Golden, CO 80401, United  States of America}
\address[label2]{Department of Mechanical Engineering, Colorado School of Mines, 1500 Illinois St, Golden, CO 80401, United  States of America}
\address[label3]{Department of Mathematics, University of North Carolina at Chapel Hill, 216 Lenoir Dr, Chapel Hill, NC 27599, United  States of America}
\address[label4]{Computational Medicine Program, University of North Carolina at Chapel Hill, 216 Lenoir Dr, Chapel Hill, NC 27599, United  States of America}

\begin{abstract}
Blood clotting involves the coupled processes of platelet aggregation and coagulation.  Simulating clotting under flow in complex geometries is challenging due to multiple temporal and spatial scales and high computational cost. \emph{clotFoam} is an open-source software developed in OpenFOAM that employs a continuum model of platelet advection, diffusion, and aggregation in a dynamic fluid environment and a simplified coagulation model with proteins that advect, diffuse, and react within the fluid and with wall-bound species through reactive boundary conditions. Our framework provides the foundation on which one can build more complex models and perform reliable simulations in almost any computational domain.
\end{abstract}


\end{frontmatter}

\section*{Current code version}
\label{sec:Current-code-version}

\begin{table}[H]
\centering
\begin{tabular}{|p{6.5cm}|p{6.5cm}|}
\hline
Current code version & 1.0 \\
\hline
Permanent link to code/repository used for this code version & \url{https://github.com/d-montgomery/clotFoam}\\
\hline
Code Ocean compute capsule & \\
\hline
Legal Code License   & GNU GPL V3 \\
\hline
Code versioning system used & git \\
\hline
Software code languages, tools, and services used & C++, C, MPI, GNU Make \\
\hline
Compilation requirements, operating environments \& dependencies & OpenFOAM-v9\\
\hline
If available Link to developer documentation/manual &  \\
\hline
Support email for questions & dmontgomery@mines.edu\\
\hline
\end{tabular}
\caption{Code metadata.}
\label{table:Code-metadata-mandatory} 
\end{table}



\section{Motivation and significance}
\label{sec:Motivation-Significance}
Blood clotting is the body's response to prevent bleeding from an injured blood vessel. The clotting process involves two main components: platelet aggregation and coagulation.  Platelet aggregation is a primarily physical process where platelets adhere to the injured vessel wall, and become activated by receptors that interact with proteins embedded in the wall. Activated platelets release agonists, such as adenosine diphosphate (ADP), which can activate and recruits more platelets to the injury where they begin to form a platelet plug. Coagulation is a biochemical process involving dozens of enzymatic reactions that occur in the fluid, on activated platelet surfaces, and on injured portions of the vessel wall. Coagulation features the interplay of positive and negative feedback loops that work collectively to encourage thrombus growth in a self-regulating manner. The reactions culminate in the generation of the enzyme thrombin on activated platelet surfaces, which is a strong platelet activation agonist, a key player in positive feedback, and converts fibrinogen into fibrin, which polymerizes and forms a stabilizing fibrin gel on the platelet plug. Platelets play a critical role in coagulation, as thrombin generation and inhibition is strongly regulated by their activated surfaces \cite{FOGELSON1998,kuharsky_surface-mediated_2001,Leiderman2011grow,miyazawa2023inhibitionB}. Thrombin generation is often used as a clinical indicator of healthy clotting, as without thrombin, a clot is typically leaky and unstable.  The interested reader can find more information about the blood clotting process and previously developed mathematical models in a number of reviews published elsewhere \cite{NeevesKeithB2016MMoH, Leidermann2018Chp,Diamond2013Review,YesudasanSumith2019Raic,AnandM.2022Cmoh}.

Spatio-temporal continuum models that employ computational fluid dynamics (CFD) have gained widespread usage to study blood clotting in devices \cite{TaylorJoshuaO.2016Doac}, in common microfluidic assays \cite{Govindarajan2018,Rojano2022}, in aneurysms \cite{BouchnitaAnass2021Mcmo} and bleeding \cite{Schoeman2016,Danes}, and to understand the effects of flow and transport on the clotting process overall \cite{Leiderman2011grow,Leiderman2012,Rezaeimoghaddam2022Cmot,Rojano2022}. A significant proportion of spatial-temporal models of clotting have been implemented using in-house codes \cite{Leiderman2011grow,Leiderman2012,Danes}, commercial tools with high licensing costs \cite{Govindarajan2018}, or are not publicly available \cite{TaylorJoshuaO.2016Doac, Wu2017, BouchnitaAnass2021Mcmo, Rojano2022}, posing a challenge for researchers with limited resources. There is growing interest in open source and freely available software tools such as OpenFOAM \cite{OpenFOAM-user-ref}. This platform provides a versatile object-oriented toolkit for developing and constructing CFD software, featuring a diverse range of solvers and discretization schemes for general grids and parallel computing.

\edit{A few in-house codes of clotting models are freely available  \cite{shankar2022three,mendez2022fibrin}, but are quite model-specific and not necessarily made for other researchers to easily build upon.} In this work we present \emph{clotFoam}, a cell-centered finite volume solver that provides a flexible framework for simulating \edit{and easily extending} a reduced model of blood clotting under flow.   The software was developed with OpenFOAM libraries and can be perceived as an extension of the transient fluid solver \emph{icoFoam}, which utilizes the PISO algorithm \cite{ISSA1986-PISO} for decoupling pressure from the fluid velocity.  \emph{clotFoam}  incorporates several additions, including a Darcy term in the fluid equations, five advection-diffusion-reaction (ADR) equations that describe platelet aggregation and the release of ADP from platelet stores upon their activation, as well as twelve ADR equations that represent a reduced model of platelet surface-mediated coagulation. The coupling of reactions among the platelet and biochemical species is managed via object-oriented programming and a modified Runga-Kutta method, employing both field and patch field operations for the computation of reactions within the fluid and on the surface of the vessel wall. The following sections of this manuscript will detail the mathematical model employed by \emph{clotFoam}, and provide a comprehensive description of the software. Two examples of the application of \emph{clotFoam} in 2-D thrombosis and 3-D hemostasis simulations are presented.  Finally, we discuss the potential implications and benefits of the \emph{clotFoam} solver for the broader modeling and simulation community.

\section{Model description}
\label{sec:Model-description-short}
The \emph{clotFoam} software is based on our previous model that used a continuum description of fluid, platelets, and platelet aggregation under flow \cite{Leiderman2011grow}. The model describes blood as an incompressible Newtonian fluid that is governed by the Navier-Stokes-Brinkman equations: 
\begin{align}
	\rho \pdv{\u}{t} + \rho (\u \cdot \grad) \u  &= - \grad \p + \mu \nabla^2\u - \mu \alpha(\theta^B)\u, \label{eq:NSB}\\
	\grad \cdot \u &= 0,
\end{align}
where $\u(\x,t)$ is the fluid velocity, $\p(\x,t)$ is pressure, $\rho$ is the fluid density, and $\mu$ is the dynamic viscosity. The Darcy term, $-\alpha(\theta^B)\u$, represents a frictional resistance to the fluid caused by a growing mass of bound platelets. The variable $\theta^B$ is the ratio of the sum of bound platelets to the maximum packing limit $\Pmax$. The permeability of the mass of bound platelets, $\alpha(\theta^B)$, decreases as $\theta^B$ increases, as it satisfies the Carman-Kozeny relation $\alpha(\theta^B) = C_{CK} (0.6\theta^B)^2/(1-0.6\theta^B)^3$, where $C_{CK} = 10^6\text{mm}^{-2}$. 

Platelets are modeled as number densities (number per volume), eliminating the need for tracking individual platelets throughout the simulation.  The dynamics of platelet aggregation are described generally by the following hindered ADR equation: 
\begin{align}
	\pdv{P^{k}}{t} &= - \div \{  W(\theta^T) ( \u P^{k} - D_P \grad P^{k} )  \} + S_k \label{eq:generalPlateletAggregation}
\end{align}
where $P^k(\x,t)$ is the $k^{th}$ platelet species and $S_k$ is a source/sink term that accounts for the transitions between different states of the platelet species.  The full system of equations in \emph{clotFoam} is in \ref{appendix:plateletAggregationModel}.
To ensure that the number of platelets at a location $\x$ does not exceed a maximum packing limit, $\Pmax$, the platelet size is considered using three phenomolgical functions: the hindered transport function $W(\theta^T)$, the adhesion region $\kadh(\x)$, and the binding affinity function $g(\eta)$.   A detailed description of these functions can also be found in \ref{appendix:plateletAggregationModel}.

The rates of activation of mobile-unactivated platelets by chemical agonists ADP and thrombin ($E_2$) are assumed to satisfy Hill functions of the form $A(c) = k_c^\text{pla} \frac{c}{c^\ast+c}$. ADP is secreted by newly activated platelets over a period of 1-5 seconds after activation. The molar concentration of ADP satisfies:
\begin{equation}
\pdv{[\ADP]}{t} = - \div \Big \{  \u [\ADP] - D_\ADP \grad [\ADP]     \Big\} + \sigma_\text{release}, \label{ADP_eqt}
\end{equation}
where $D_\text{ADP}$ is the diffusion coefficient, and the source term is defined as: 
\begin{align}
\sigma_\text{release} (\x, t)&= \int_0^\infty \hat A R(\tau) \pdv{t} ( P^{b,a} + P^{se,a} )(\x,t-\tau) d\tau . \label{eq:ADP_source}
\end{align}
$\hat A$ is the total concentration of ADP released by an activated platelet, and $\hat A R(\tau) $ is the rate of release of ADP $\tau$ seconds after activation. The rate function, $R(\tau)$, utilized by \emph{clotFoam} is similar to the one in our previous work \cite{Leiderman2011grow}, but here uses a bell curve centered at 3 seconds, $R(\tau) = \frac{1}{\sqrt{\pi}}\exp(-(\tau - 3)^2)$, and is normalized such that $\int_0^\infty R(\tau) \, d\tau = 1. $ 

\begin{figure}[H]
\makebox[\linewidth][c]{%
  \includegraphics[width=\textwidth,trim={8cm 6cm 8cm 9cm},clip]{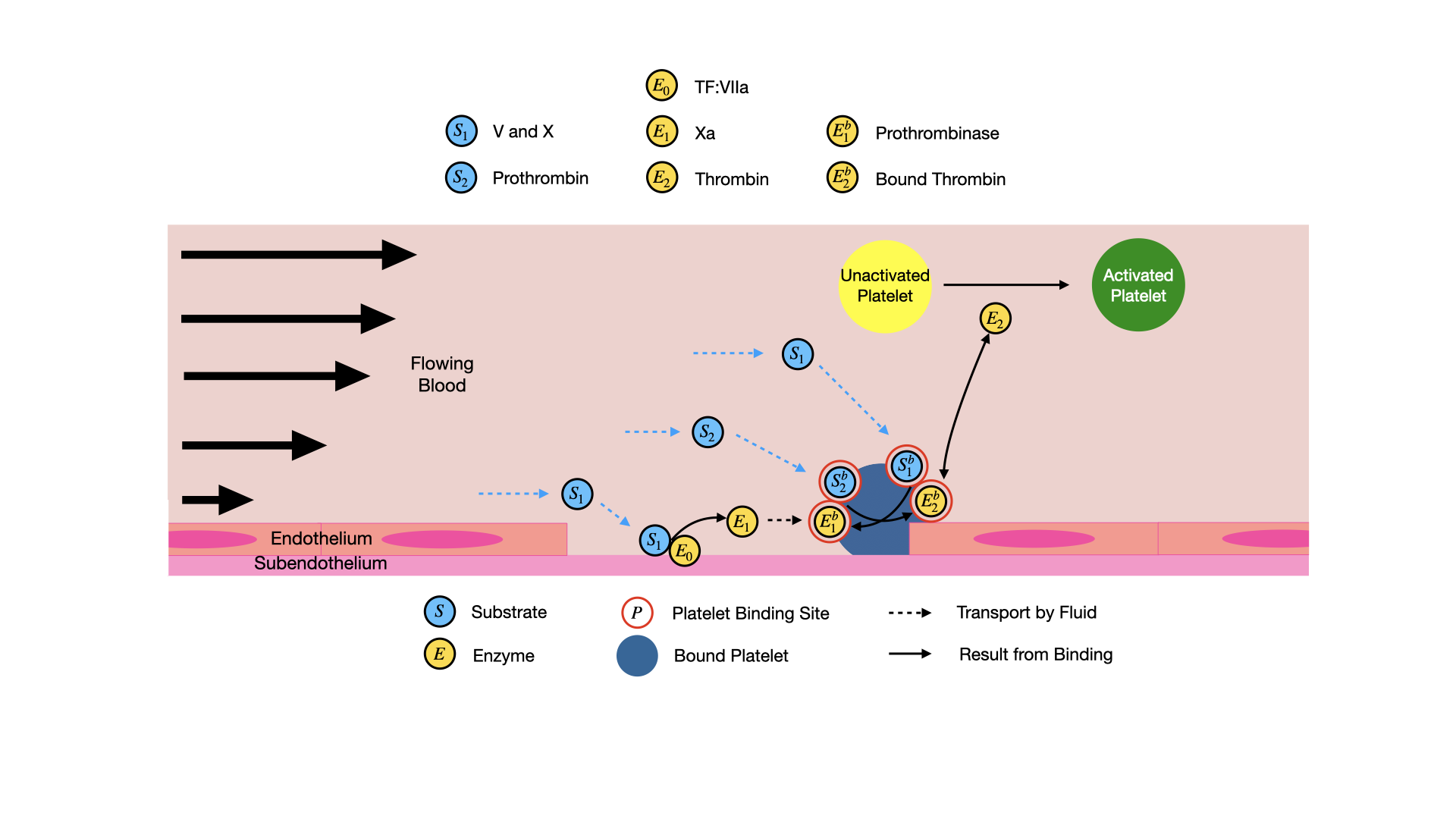}
}
\caption{Schematic representation of the reduced model of thrombin generation with positive feedback. The light blue circles represent a substrate, yellow circles denote enzymes, and the red circles represent binding sites on the platelet surface. Dotted lines indicate transport by the fluid, while the solid lines represent binding interactions. \textit{Note: Platelets have diameters near 3 $\mu$m, while endothelial cells are typically 50-70 $\mu$m in length and 0.1-10 $\mu$m in thickness \cite{feletou2011endothelium}.}}
\label{fig:coagCascade}
\end{figure}

The reduced coagulation model consists of various biochemical species in molar concentrations and categorized as fluid-phase, platelet-bound, or subendothelium-bound. In an effort to enhance the adaptability of \emph{clotFoam}, the number of biochemical species has been reduced from 50 \cite{Leiderman2011grow,Leiderman2012} to 12. This model is an extension of a previously published ODE model \cite{FOGELSON1998} that includes positive feedback and enzyme inhibition, two features necessary to capture the bursting thrombin behavior observed in coagulation. We extended the ODE model to a PDE model where reactions occur on two surfaces (subendothelium and activated platelets) instead of one, and the species are subjected to flow. The model is detailed in our \ref{appendix:coagulationModel}, and a schematic is presented in Figure \ref{fig:coagCascade}.  The reduced model is summarized as follows where $S$ denotes a substrate or zymogen, and $E$ denotes an enzyme:

\begin{enumerate}
    \item Fluid-phase substrate, $S_1$, comes into contact with enzyme, $E_0$, bound to the subendothelium. $E_0$ converts the substrate to an enzyme, $E_1$.
    \item $E_1$ binds to the activated platelet surface and becomes the bound species, $E_1^b$.
    \item Additional fluid-phase substrates, $S_2$, bind to the platelet surface and become $S_2^b$. Upon activation by $E_1^b$, the platelet-bound substrates, $S_2^b$, are converted into a second platelet-bound enzyme, $E_2^b$, which we consider to be similar to thrombin.
    \item $E_2^b$ activates platelet-bound substrates, forming more enzymes in a positive feedback loop.
    \item $E_2$ activates mobile unactivated platelets. 
    
\end{enumerate}

\section{Software description}
\label{sec:Software-description}
\begin{figure}[H]
\makebox[\linewidth][c]{%
  \includegraphics[width=0.6\textwidth,]{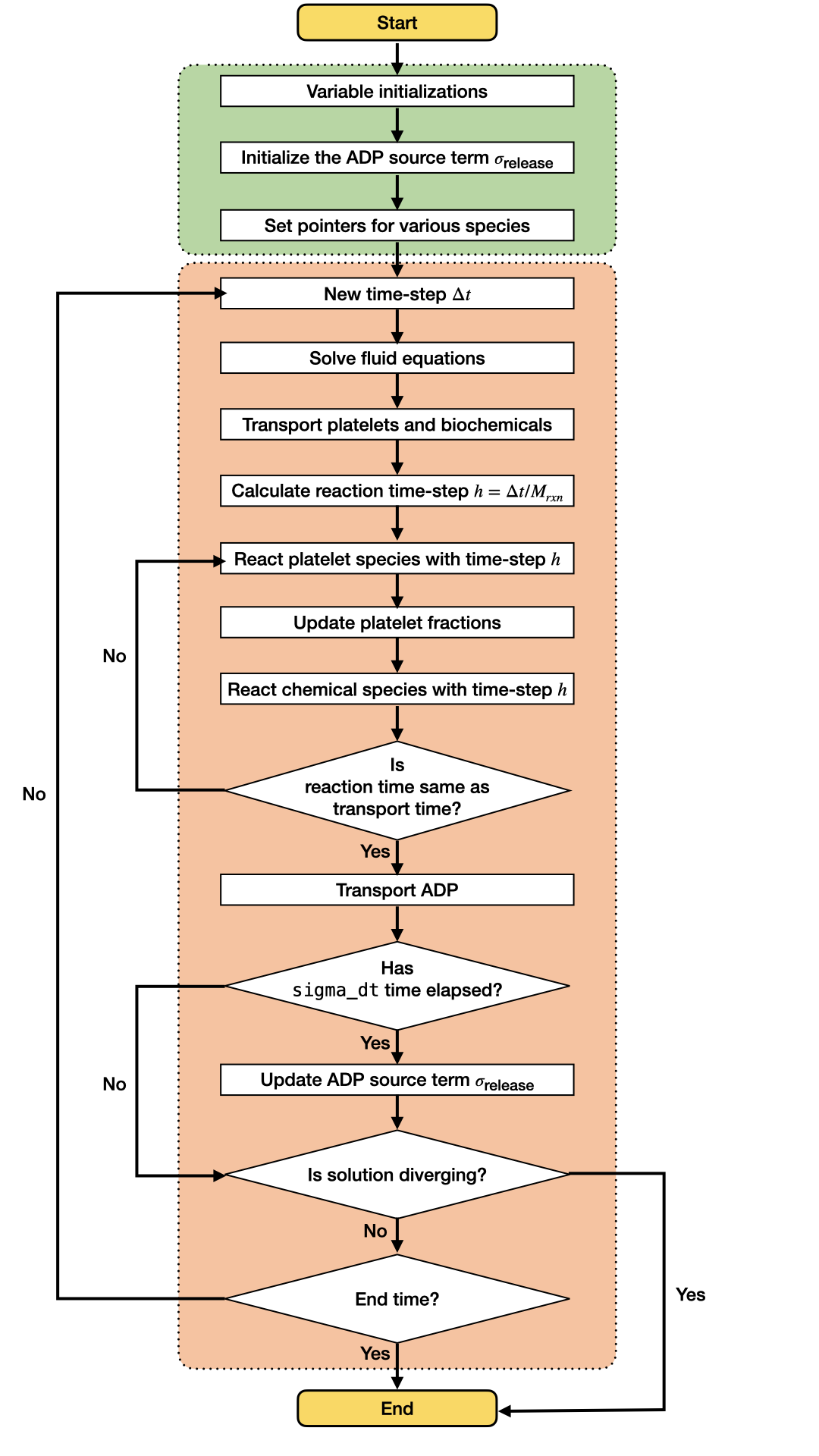}
}
\caption{Flow chart of platelet-mediated coagulation solver within OpenFOAM.}
\label{fig:softwareFlowChart}
\end{figure}

\emph{clotFoam} is an open-source software distributed under the GNU General Public License, compiled using the OpenFOAM-v9 libraries. The software can be compiled on any system where the OpenFOAM-v9 libraries are installed, and full installation instructions are provided in the repository. The code is written in C++ and can be adapted to simulate clotting in a wide variety of domains with few limitations to the mesh. The use of object-oriented programming enables the management of platelets and biochemical species as objects, which simplifies the implementation of more complex coagulation and clotting models.  The repository contains two illustrative examples, with the expectation that its range will expand through community contributions and author updates as the framework develops further.  \emph{clotFoam} is fully parallelizable for high performance computing (HPC) using the message passing interface (MPI) framework.  The solution algorithm employed by \emph{clotFoam} is illustrated in Figure \ref{fig:softwareFlowChart} and is described in more detail in the following subsections.

\subsection{Mesh requirements}
\begin{figure}[H]
\captionsetup[subfigure]{justification=Centering}
\makebox[\linewidth][c]{%
\begin{subfigure}[t]{\textwidth}
\centering
\hspace*{-1.02cm}  
    \includegraphics[width=0.675\textwidth,trim={3cm 9cm 4cm 9cm},clip]{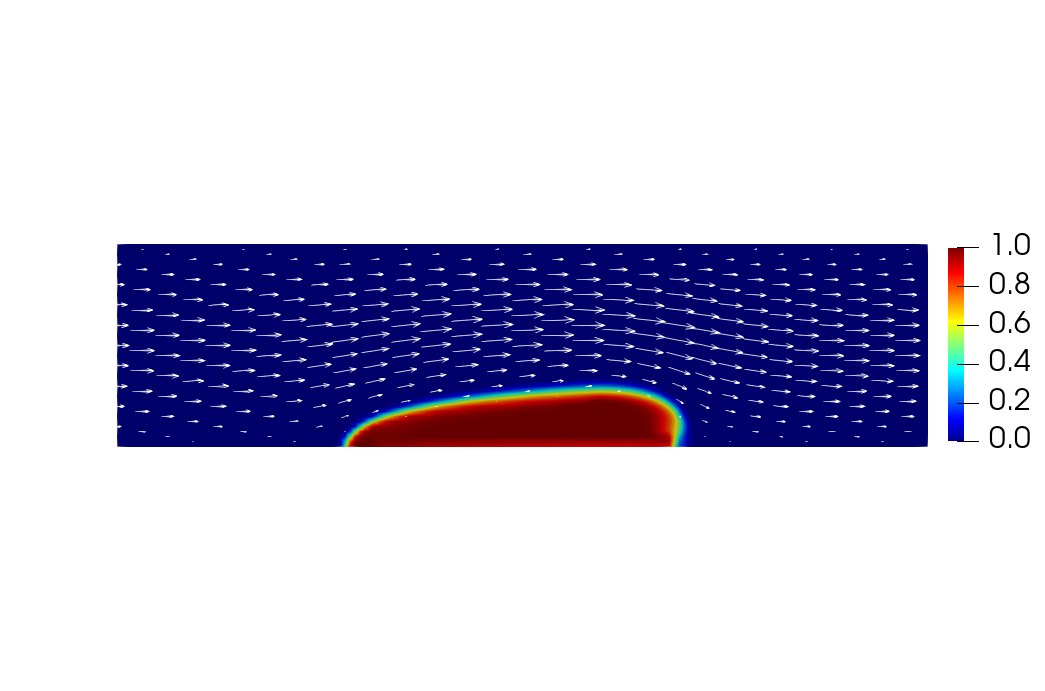}
\end{subfigure} 
}

\medskip 
\makebox[\linewidth][c]{%
\begin{subfigure}[t]{\textwidth}
\centering
    \includegraphics[width=0.72\textwidth,trim={10cm 12cm 10cm 12.75cm},clip]{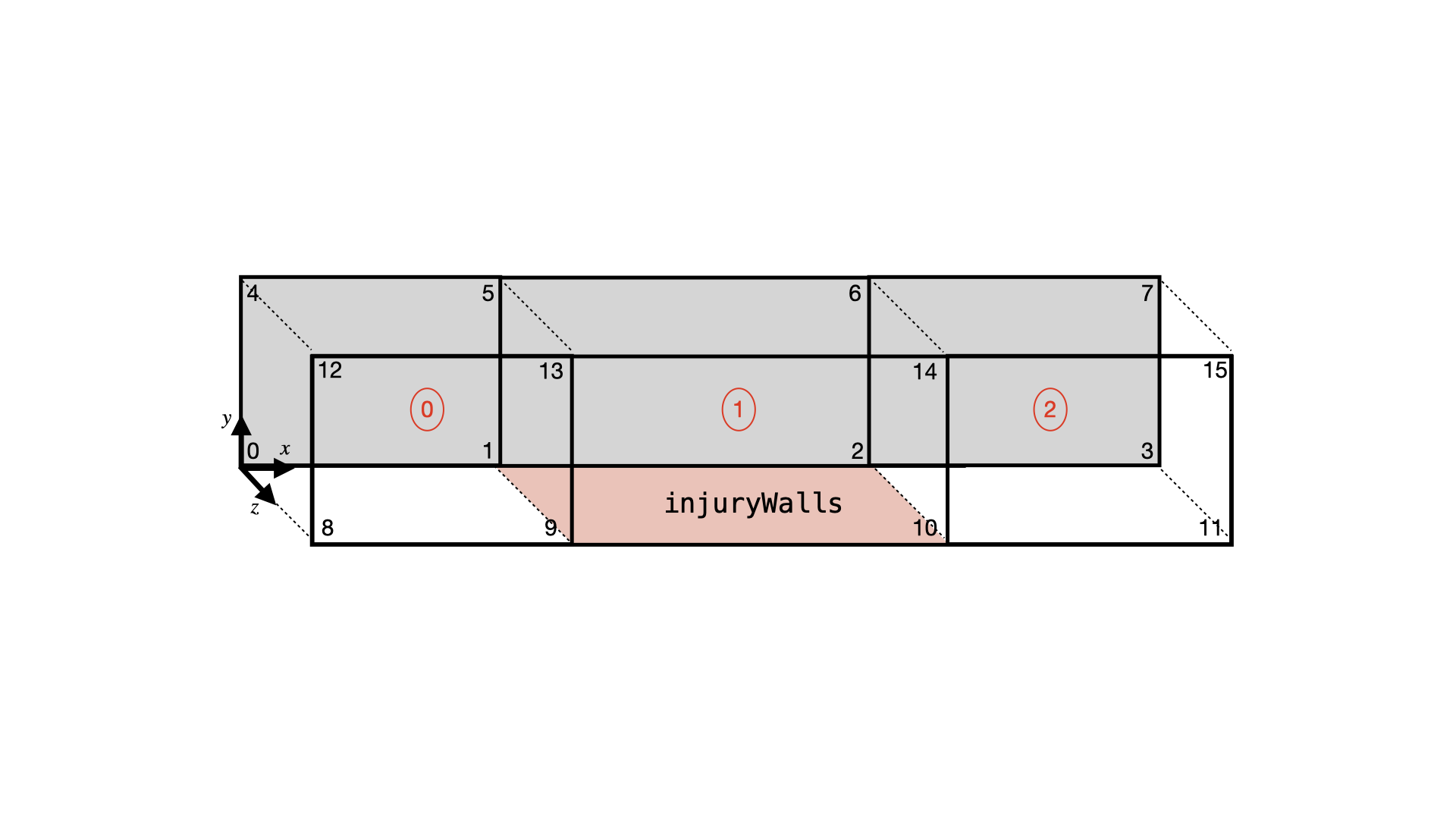}
\end{subfigure}
}%
\caption{Example of thrombus growth in a corresponding domain defined with \texttt{blockMesh} using three blocks and 16 vertices. The \texttt{injuryWalls} patch is defined by vertices \{1, 9, 10, 2\}, which is used by \emph{clotFoam} to determine the location of the reactive boundary conditions in the coagulation model.}
\label{fig:injuryBlock_blockMesh}
\end{figure}

To properly define the injury region, the mesh needs to satisfy two requirements. First, the reactive boundary conditions and subendothelium-bound species are defined exclusively on a patch known as "\texttt{injuryWalls}". This patch is constructed to be the wall of an injury block within the domain using the \texttt{blockMesh} tool, as depicted in Figure \ref{fig:injuryBlock_blockMesh}. Alternatively, users can use the \texttt{topoSet} tool to define the \texttt{injuryWalls} patch.  Second, the mesh at the injury site must feature cell widths, heights, and depths no greater than a platelet diameter $\Pdiam$, which has a default value of 3 $\mu$m. This requirement is a consequence of how the adhesion region $H_\text{adh}(\x)$ is defined.

\subsection{Numerical methods} \label{sec:numericalMethods}
\edit{
The fluids solver in \emph{clotFoam} is built upon OpenFOAM's transient fluid solver \emph{icoFoam}, which implements the PISO algorithm as a predictor and corrector method. In each time step, the momentum equation is solved once, followed by multiple pressure and velocity corrections. During the discretization of the Navier-Stokes-Brinkman equations \eqref{eq:NSB}, the Darcy term is treated implicitly as a source term. This treatment involves utilizing the bound platelet fraction, $\theta^B$, from the previous time step. By incorporating the bound platelet fraction in this manner, we not only enhance the stability of the fluids solver but also ensure that the pressure corrections in the PISO algorithm are influenced by the presence of the porous media. Consequently, we are able to accurately capture the influence of the porous media on the fluid flow, resulting in simulations that exhibit improved reliability and robustness. Further details regarding other discretization schemes employed in this work can be found in \ref{appendix:discretization}.
}

The model accounts for the transport of both platelet and biochemical species with equations that incorporate advection, diffusion, and reactions (ADR) with other species. The reactions occur on a smaller time scale than the transport processes, specifically, the software's default reaction time-step is half of the transport time-step. As such, the software employs a fractional-step method to decouple the transport from the reaction terms in the equations. Thus, the reaction equations are solved multiple times during each fluid time-step. The general ADR equation for each species is:
\begin{equation}
    \pdv{C_i}{t} = - \div (\v C_i - D_c \grad C_i) + R_i(C_1, C_2, ...,C_n), \label{eq:generalADR}
\end{equation}
where $C_i$ is the $i^{th}$ species with $i = 1, \dots, n$, $\v$ is a fluid velocity (not necessarily $\u$ from \eqref{eq:NSB}), $D_c$ is the diffusion coefficient, and $R_i$ is a reaction term that can depend on multiple species.  

A two-step fractional-step method is used to march equation \eqref{eq:generalADR} forward in time for each time step:
\begin{enumerate}
    \item  Solve the transport equation with a temporal step size $\Delta t$: 
    \begin{equation}
    \pdv{C_i}{t} = - \div (\v C_i - D_c \grad C_i).
    \end{equation}
    \item Update the solution by solving the coupled reaction equations \texttt{M\_rxn} times with a temporal step size $h = \Delta t/ M_\text{rxn}$:
    \begin{equation}
        \pdv{C_i}{t} = R_i(C_1, C_2, \dots, C_n).
    \end{equation}
\end{enumerate}
\edit{
The transport equations are discretized using the finite volume method (FVM) as discussed in Appendix \ref{appendix:discretization}, 
}
 while the reaction equations are solved using a coupled fourth-order Runge-Kutta (RK4) method.  The parameter \texttt{M\_rxn} determines how many times the reaction equations are solved per fluid time-step, and is specified in the \linebreak \texttt{\$FOAM\_CASE/constant/inputParameters} file. The default value is $\texttt{M\_rxn}=2$, however, it should be noted that this parameter is dependent on the specific problem and may need to be adjusted for flows with higher wall-shear rates.

In the mobile platelet equations described in \ref{appendix:plateletAggregationModel}, the flux vector $\va{j} = \u P - D_P \grad P$ is scaled by a hindered transport function $W(\theta^T)$ to limit the transport of platelets near the growing thrombus. Prior to FVM discretization of the advective and diffusive fluxes, the total platelet fraction $\theta^T$ must be interpolated to the cell faces. The choice of interpolation method is determined by the mechanism of transport. By implementing a combination of interpolation schemes, the flux of platelets into a spatial location is effectively constrained, ensuring that the maximum value of the sum of all platelet species at a spatial location remains below or equal to a maximum packing density $\Pmax$. To interpolate the total platelet fraction for the advective flux, $W(\theta^T) \u P$, a \texttt{downwind} scheme is used with respect to the fluid velocity $\u$. This is because the fluid velocity is assumed to only be hindered by a thrombus that is downstream. Conversely, for the diffusive flux, $W(\theta^T)D_P \grad P$, the total platelet fraction is interpolated using a \texttt{localMax} scheme, as the diffusion rate within the thrombus is expected to be smaller than the rate outside of the thrombus. 

The ADP equation \eqref{ADP_eqt} is not solved using the fractional-step method, because the source term is updated infrequently.   When platelets become bound, they release ADP into the fluid for up to 6 seconds.  The secretion of ADP is modeled by the source term, $\sigma_\text{release}$, as defined in equation \eqref{eq:ADP_source}, and can be restricted to the interval $\tau \in [0,\, 6]$ due to the bell-shaped distribution of $R(\tau)$. However, the computation of $\sigma_\text{release}$ is memory intensive as the number of newly bound platelets $\pdv{t} ( P^{b,a} + P^{se,a} )(\x,t_n-\tau)$ must be stored in memory for up to 6 seconds.  To reduce computational cost, a coarse discretization of $\tau$ is employed to calculate $\sigma_\text{release}$. The number of newly bound platelets are computed and stored at a user-specified interval $\Delta \tau$.  The discretization of $\sigma_\text{release}$ at time $t = t_n$ is then implemented using the trapezoid rule:
{\footnotesize
\begin{align}
\sigma_\text{release}(\x, t_n) &= \int_0^{\tau_{f}} \hat A R(t') \pdv{t} ( P^{b,a} + P^{se,a} )(\x,t_n-t') \, dt', \nonumber \\[5pt]
&= \int_{t_n - \tau_f}^{t_n} \hat A R(t_n-\tau) \pdv{t} ( P^{b,a} + P^{se,a} )(\x,\tau) \, d\tau, \hspace{5mm}(\text{by substituting $\tau = t_n - t'$})\nonumber\\[5pt]
&\approx \sum_{k= 0}^{N_\tau - 1} \hat A \, \frac{\Delta \tau}{2}    \bigg\{  R(t_n - \tau_k) 
\pdv{t} ( P^{b,a} + P^{se,a} )(\x,\tau_k ) + R(t_n - \tau_{k+1}) \pdv{t} ( P^{b,a} + P^{se,a} )(\x,\tau_{k+1} ) \bigg\}, \label{sigma_disc} 
\end{align}
}%
where $\tau_f$ and $\Delta \tau$ are defined as \texttt{sigma\_Tf} and \texttt{sigma\_dt} respectively in the \texttt{inputParameters} dictionary. Lastly, the number of newly bound platelets is approximated as:
\begin{align}
	\pdv{t} ( P^{b,a} + P^{se,a} )(\x,t_n) &\approx  \Pmax \frac{\theta^B_{n} - \theta^B_{n-1}}{t_n - t_{n-1}}.
\end{align}

\subsection{Managing platelet and biochemical species with polymorphism}
\begin{figure}[H]
\makebox[\linewidth][c]{%
  \includegraphics[width=\textwidth,trim={0cm 3cm 0cm 0cm},clip]{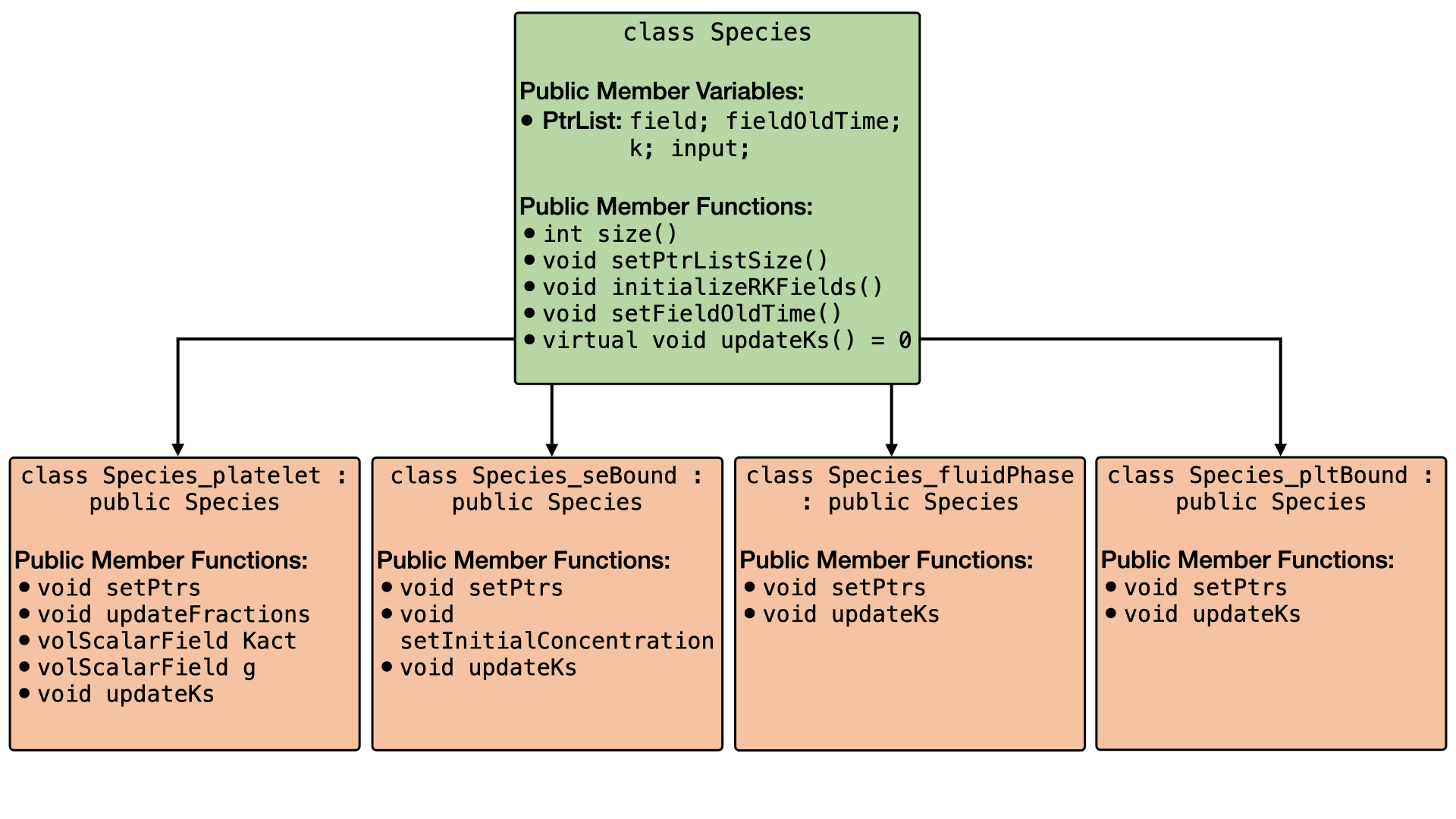}
}
\caption{Graph of the \texttt{Species} class and derived classes \texttt{Species\_platelet, Species\_seBound, Species\_fluidPhase, Species\_pltBound}.  All derived classes inherit the public member variables and public member functions defined in the \texttt{Species} class.}
\label{fig:SpeciesClassGraph}
\end{figure}

Models of blood clotting typically involve multiple platelet and biochemical species. For instance, the Leiderman-Fogelson model \cite{Leiderman2011grow} consists of four platelet species and 50 biochemical species. To address this complexity, \emph{clotFoam} has been developed to accommodate models with any number of platelet and biochemical species. 
These models are implemented following a polymorphic approach, with an
abstract base class called \texttt{Species} from which four classes are derived to align with the species defined in the mathematical model: \texttt{Species\_platelet, Species\_seBound, Species\_fluidPhase, Species\_pltBound}. The inheritance relationship of these classes is depicted in Figure \ref{fig:SpeciesClassGraph}.

The \texttt{Species} object consists of four pointer lists to facilitate the management of the subspecies fields, including field values (solutions), previous field values (solutions from intermediate time steps used in the fractional-step method and RK4 solver), $k$ values (computed for the RK4 solver), and the argument of the $k$ values (inputs for the $k$ values in the RK4 solver). The public member functions of the derived classes enable the setting of pointers to other \texttt{Species} objects and define the reaction functions specific to their corresponding subspecies. The function \texttt{updateKs} computes and retains the reaction term, $R_i$, for each subspecies, utilizing input parameters passed from the RK4 method. The initialization of every derived \texttt{Species} object occurs within the \texttt{createFields.H} file, and the pointers are set in \texttt{setSpeciesPointers.H}.

\subsection{Adapting the framework for different clotting models}
\emph{clotFoam} was developed with the aim of promoting scientific research and reproducibility in the field of hemostasis, thrombosis, and clotting disorders. Thus, we illustrate the procedure for introducing a new species into the software. Consider the addition of a fluid-phase species $I$, that inhibits thrombin indefinitely.  This reaction can be written as:
\begin{equation}
    I + E_2 \xrightarrow{k_I} E_{2,\text{inh}},
\end{equation}
where $k_I$ is the association rate of inhibitor and enzyme, and $E_{2,\text{inh}}$ is the resulting inhibited thrombin.  
Although the addition of two new species, $I$ and $E_{2,\text{inh}}$, is necessary for incorporating the desired reaction into the software, we will focus our discussion on the steps involved in including just the inhibitor species as it applies to any additional species. The new reaction term that is added to the ADR equation for $I$ is:
\begin{equation}
    \text{reaction rate:} -k_I I E_2, \label{eq:sampleReaction}
\end{equation}
which fits the form of equation \eqref{eq:generalADR}, with a reaction term defined by the right-hand side.  The following sequential steps outline the procedure for adding $I$ to the software:

\begin{enumerate}
    \item Within the case directory \texttt{\$FOAM\_CASE}:
    \begin{enumerate}
        \item In the \texttt{0} directory:
        \begin{enumerate}
            \item Create a new field, such as \texttt{fluidPhase\_I}.
        \end{enumerate}
        \item In the \texttt{constant/inputParameters} file:
        \begin{enumerate}
            \item Update the number of fluid-phase species, \texttt{num\_fluidPhase}.
            \item Incorporate the parameter \texttt{kI} using appropriate units.
        \end{enumerate}
    \end{enumerate}
    \item Within the \emph{clotFoam} software:
    \begin{enumerate}
        \item In the \texttt{chemConstants.H} file:
        \begin{enumerate}
            \item Read in the parameter \texttt{kI} defined in the \texttt{inputParameters} dictionary.
        \end{enumerate}
        \item In the \texttt{Species\_fluidPhase.H} class file:
        \begin{enumerate}
            \item Update the reaction function \texttt{updateKs} to include 

            \vspace{-1cm}
            \begin{center}
            \[\texttt{k[4] = -kI*I*E2 },\]
            \end{center}
            where \texttt{I} is a reference to the input necessary for the RK4 solver, 
            
            \begin{center}
             \texttt{const volScalarField\& I = input[4]},
             \end{center}
            
             defined at the beginning of the \texttt{updateKs} function. 
            \item If necessary, update the pointers in the constructor and in the \texttt{setPtrs} function (not required for this example). \label{item:pointers}
        \end{enumerate}
        \item In the \texttt{createFields.H} file:
        \begin{enumerate}
            \item Add the \texttt{fluidPhase\_I} field to the 4$^{th}$ index of the \texttt{PtrList} called \linebreak \texttt{fluidPhase.field}. 
        \end{enumerate}
        \item In the \texttt{setSpeciesPointers.H} file:
        \begin{enumerate}
            \item If required, update the \texttt{fluidPhase.setPtrs} argument corresponding to step \ref{item:pointers}.
        \end{enumerate}
    \end{enumerate}
\end{enumerate}

\section{Illustrative Examples}
\label{sec:Illustrative-Examples}
Here, we present two illustrative examples for which the code and instructions are included in the repository. These examples showcase the capabilities of the \emph{clotFoam} solver in simulating blood clotting phenomena. The first example is a 2D thrombosis case based on previous results published by our group \cite{Leiderman2011grow,Leiderman2012}. The second example is a 3D simulation of hemostasis that replicates the microfluidic device described in our previous study by Schoeman et al. \cite{Schoeman2016}.  \edit{Additional examples of convergence and validation can be found in the supplementary material.}

\subsection{Thrombosis in a rectangular channel} \label{sec:Illustrative-Example-Thrombosis}
\begin{figure}[H]
\captionsetup[subfigure]{justification=Centering}

\makebox[\linewidth][c]{%
\begin{subfigure}[t]{0.33\textwidth}
\hspace*{0.35cm}  
\centering
    \includegraphics[width=\textwidth,trim={5cm 9cm 5cm 9cm},clip]{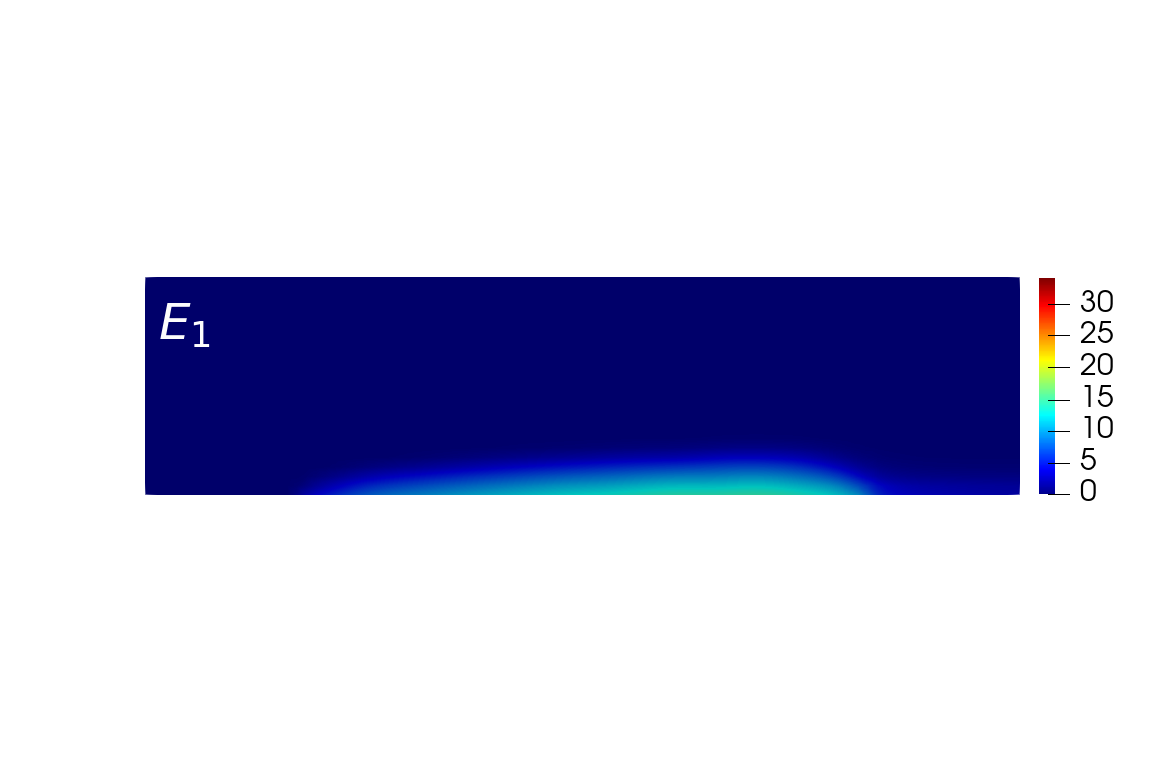}
\end{subfigure}
\begin{subfigure}[t]{0.33\textwidth}
\hspace*{0.35cm}  
\centering
    \includegraphics[width=\textwidth,trim={5cm 9cm 5cm 9cm},clip]{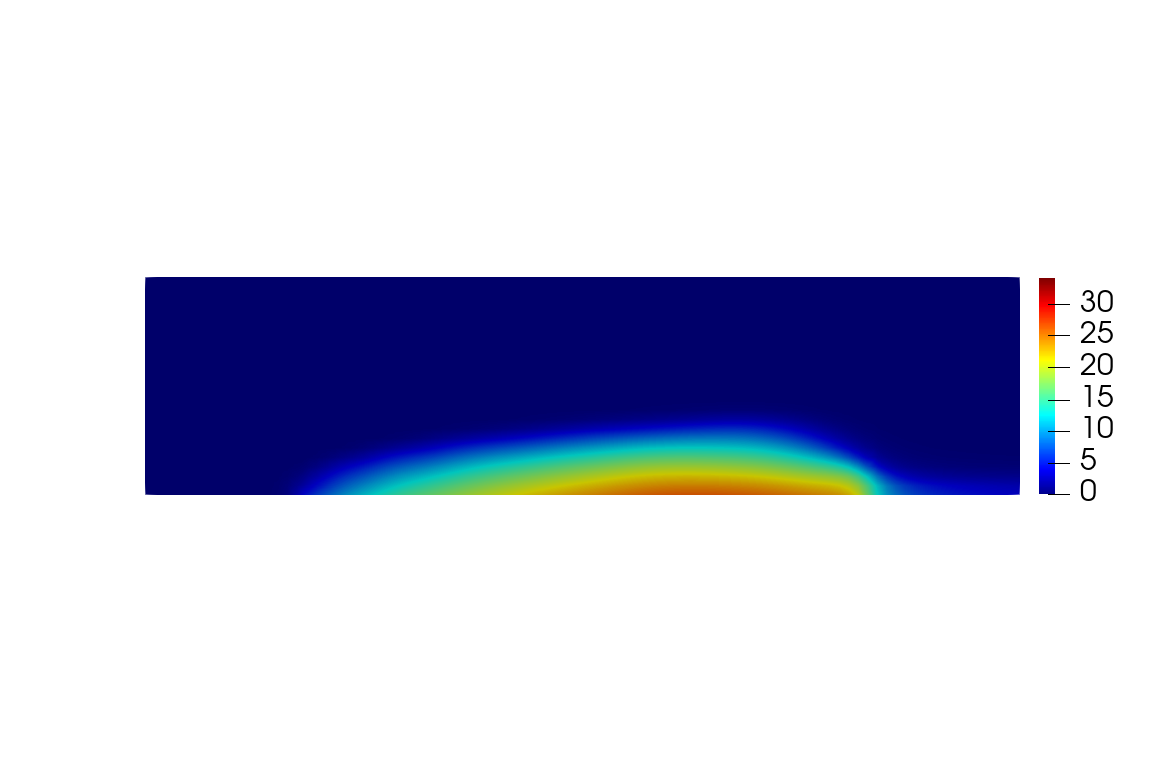}
\end{subfigure}
\begin{subfigure}[t]{0.38\textwidth}
\hspace*{0.35cm}  
\centering
    \includegraphics[width=\textwidth,trim={5cm 9cm 0cm 9cm},clip]{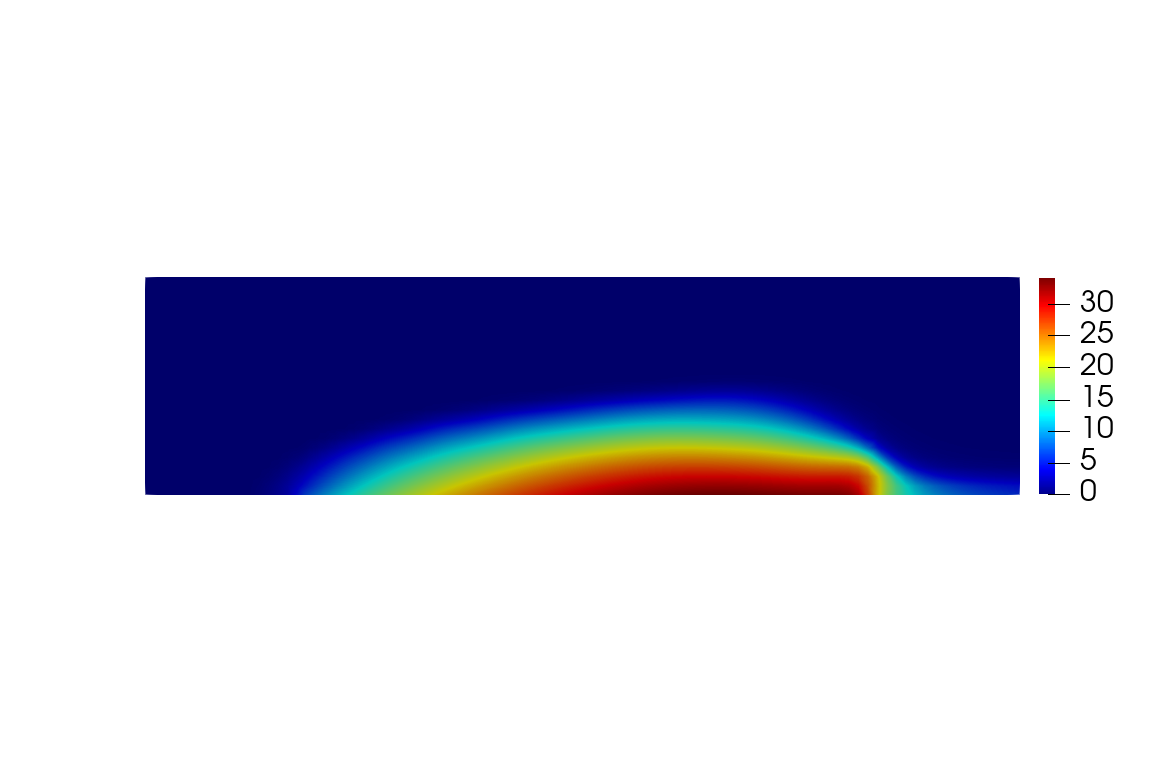}
\end{subfigure}
}

\makebox[\linewidth][c]{%
\begin{subfigure}[t]{0.33\textwidth}
\hspace*{0.35cm}  
\centering
    \includegraphics[width=\textwidth,trim={5cm 9cm 5cm 9cm},clip]{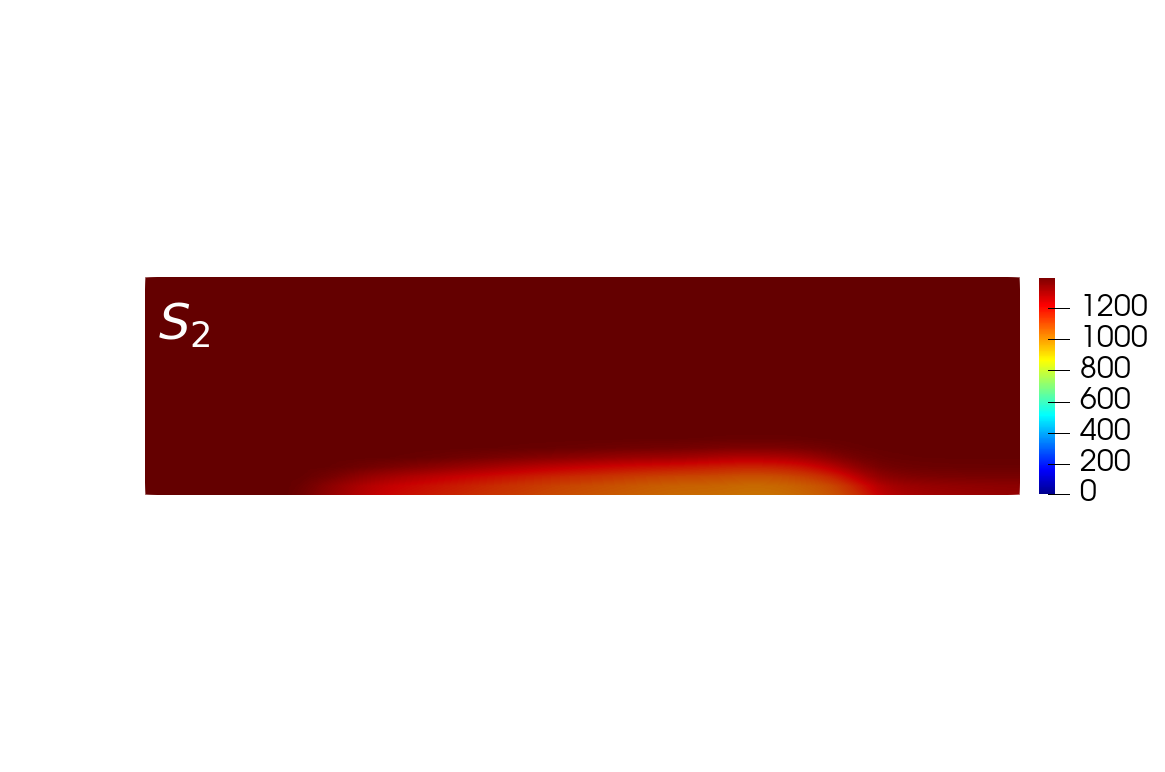}
\end{subfigure}
\begin{subfigure}[t]{0.33\textwidth}
\hspace*{0.35cm}  
\centering
    \includegraphics[width=\textwidth,trim={5cm 9cm 5cm 9cm},clip]{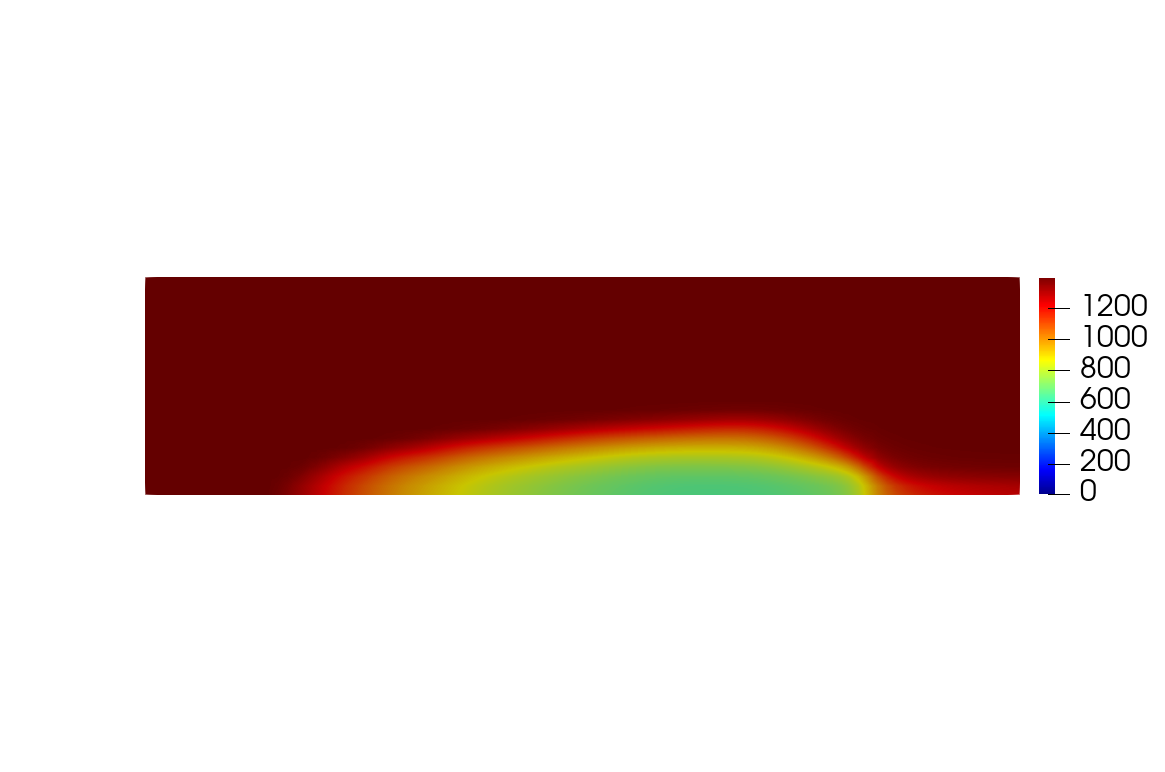}
\end{subfigure}
\begin{subfigure}[t]{0.38\textwidth}
\hspace*{0.35cm}  
\centering
    \includegraphics[width=\textwidth,trim={5cm 9cm 0cm 9cm},clip]{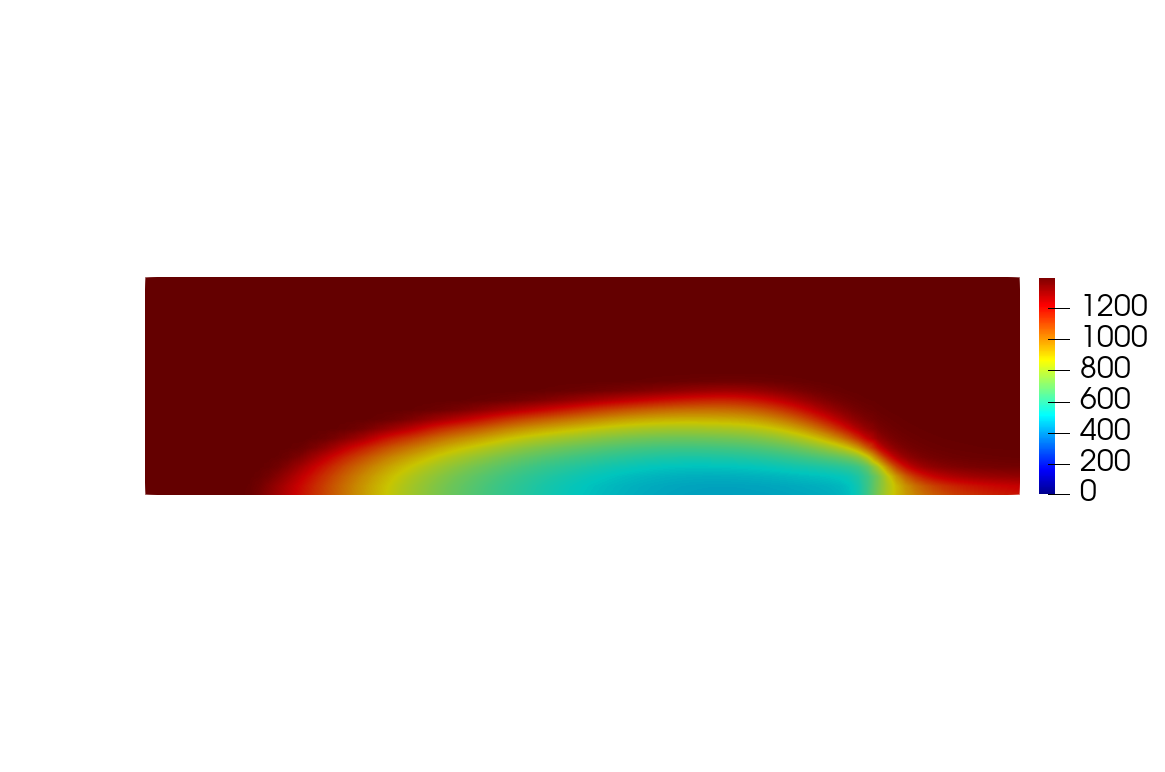}
\end{subfigure}
}

\makebox[\linewidth][c]{%
\begin{subfigure}[t]{0.33\textwidth}
\hspace*{0.35cm}  
\centering
    \includegraphics[width=\textwidth,trim={5cm 9cm 5cm 9cm},clip]{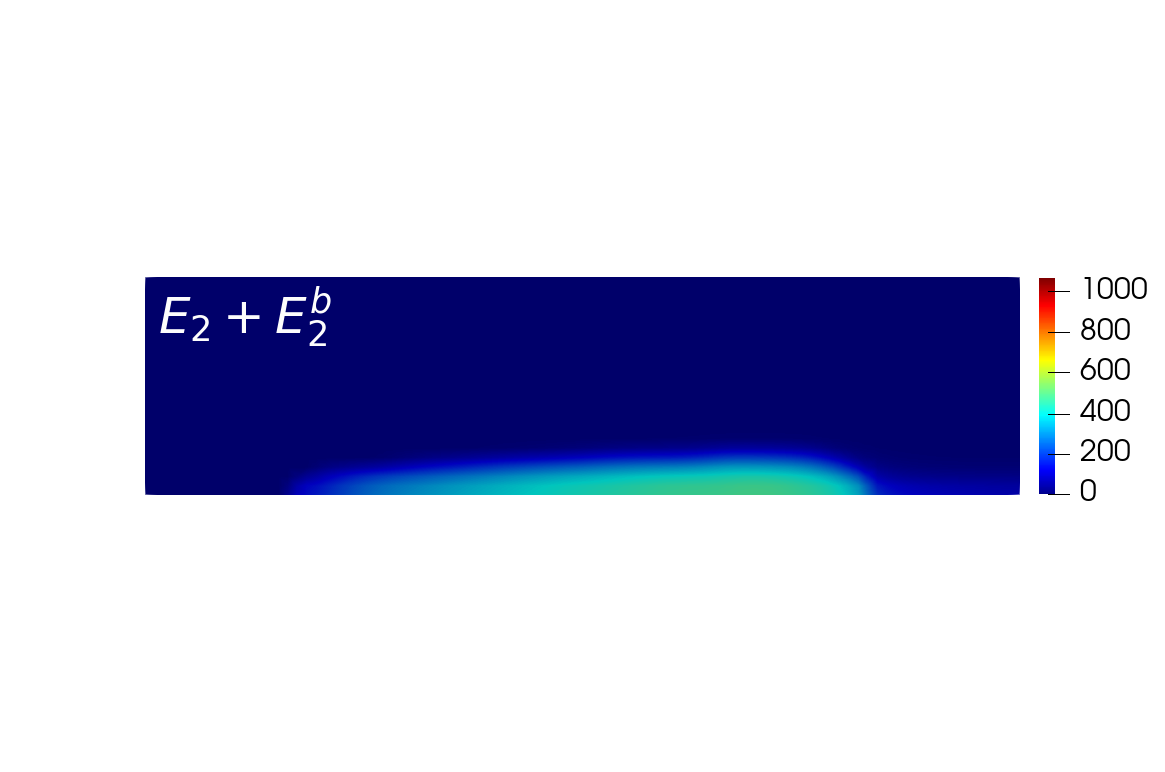}
\end{subfigure}
\begin{subfigure}[t]{0.33\textwidth}
\hspace*{0.35cm}  
\centering
    \includegraphics[width=\textwidth,trim={5cm 9cm 5cm 9cm},clip]{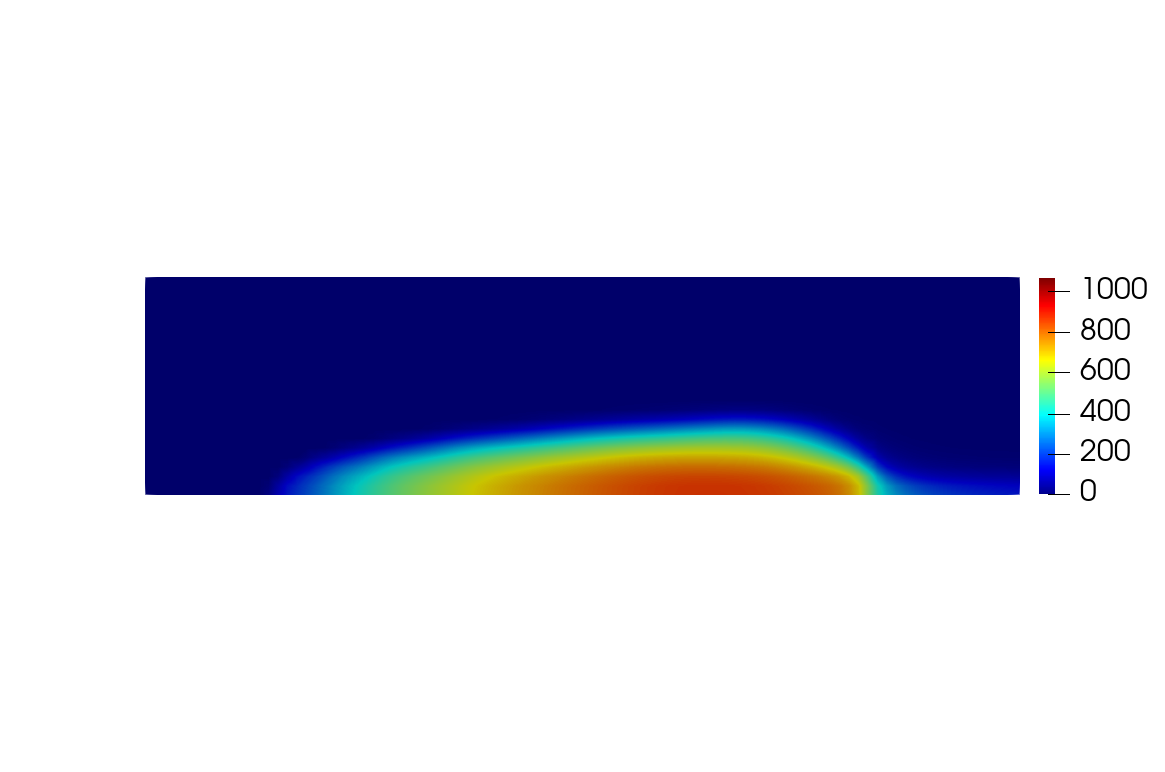}
\end{subfigure}
\begin{subfigure}[t]{0.38\textwidth}
\hspace*{0.35cm}  
\centering
    \includegraphics[width=\textwidth,trim={5cm 9cm 0cm 9cm},clip]{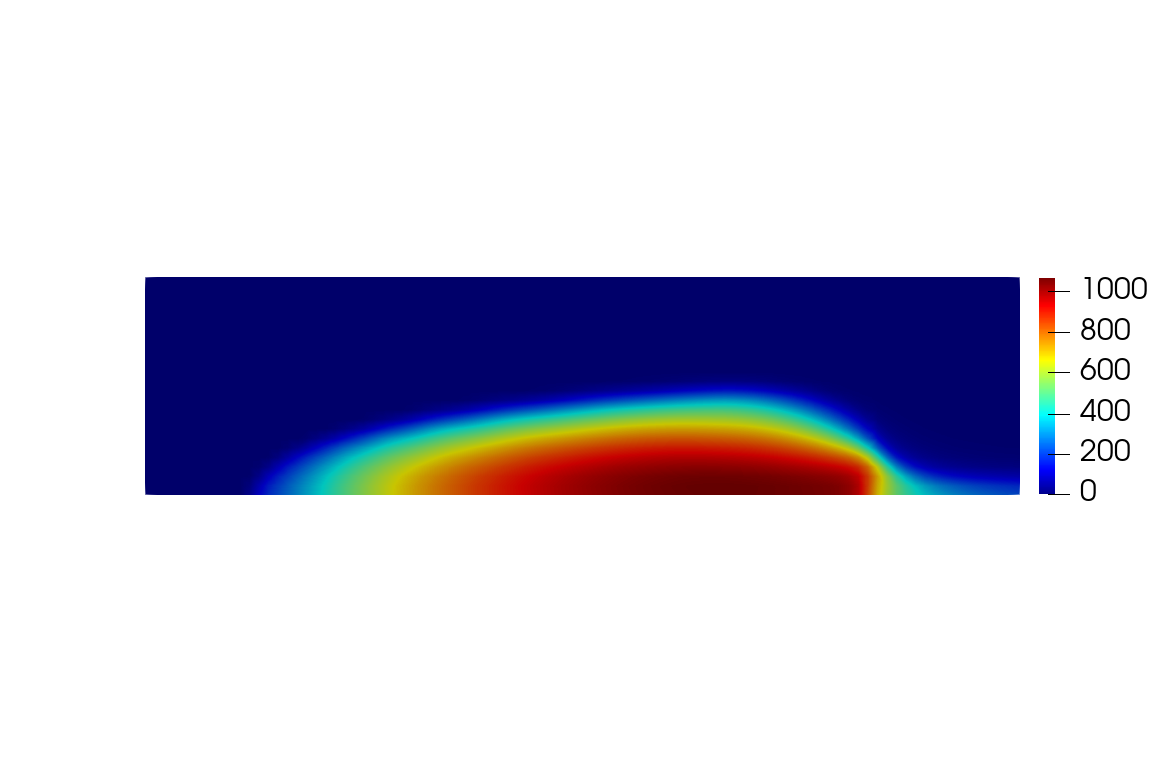}
\end{subfigure}
}

\makebox[\linewidth][c]{%
\begin{subfigure}[t]{0.33\textwidth}
\hspace*{0.35cm}  
\centering
    \includegraphics[width=\textwidth,trim={5cm 9cm 5cm 9cm},clip]{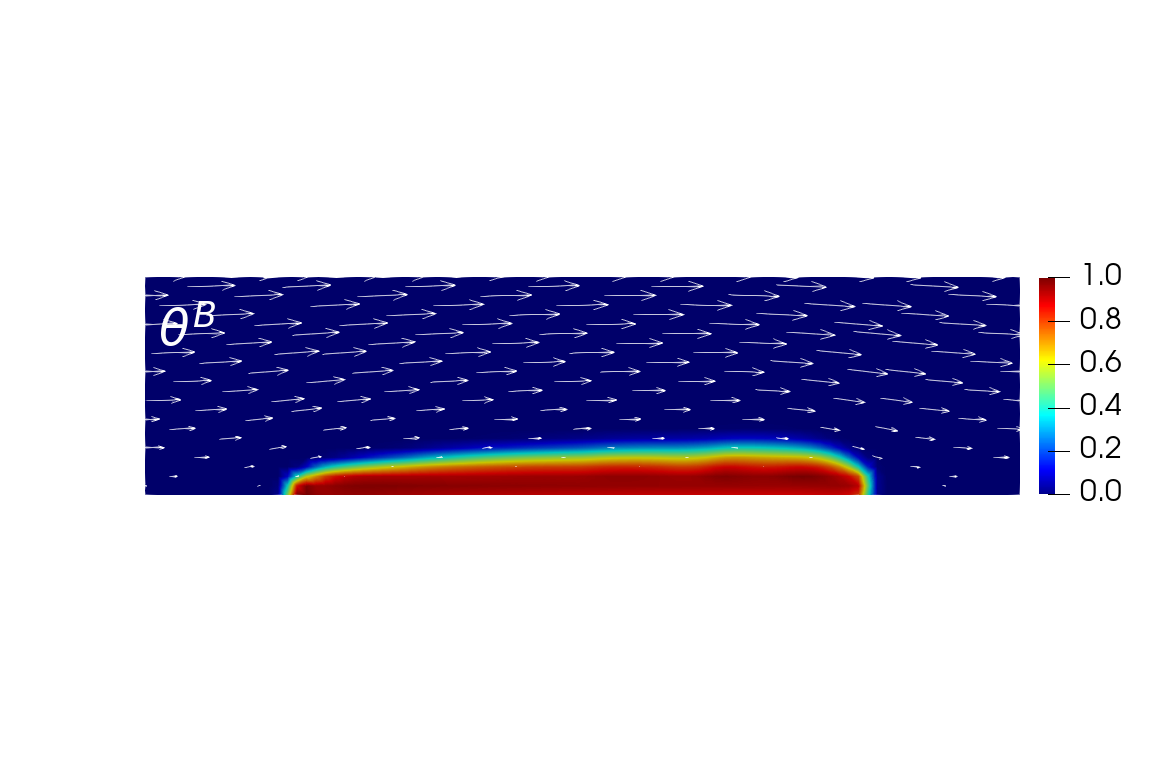}
\end{subfigure}
\begin{subfigure}[t]{0.33\textwidth}
\hspace*{0.35cm}  
\centering
    \includegraphics[width=\textwidth,trim={5cm 9cm 5cm 9cm},clip]{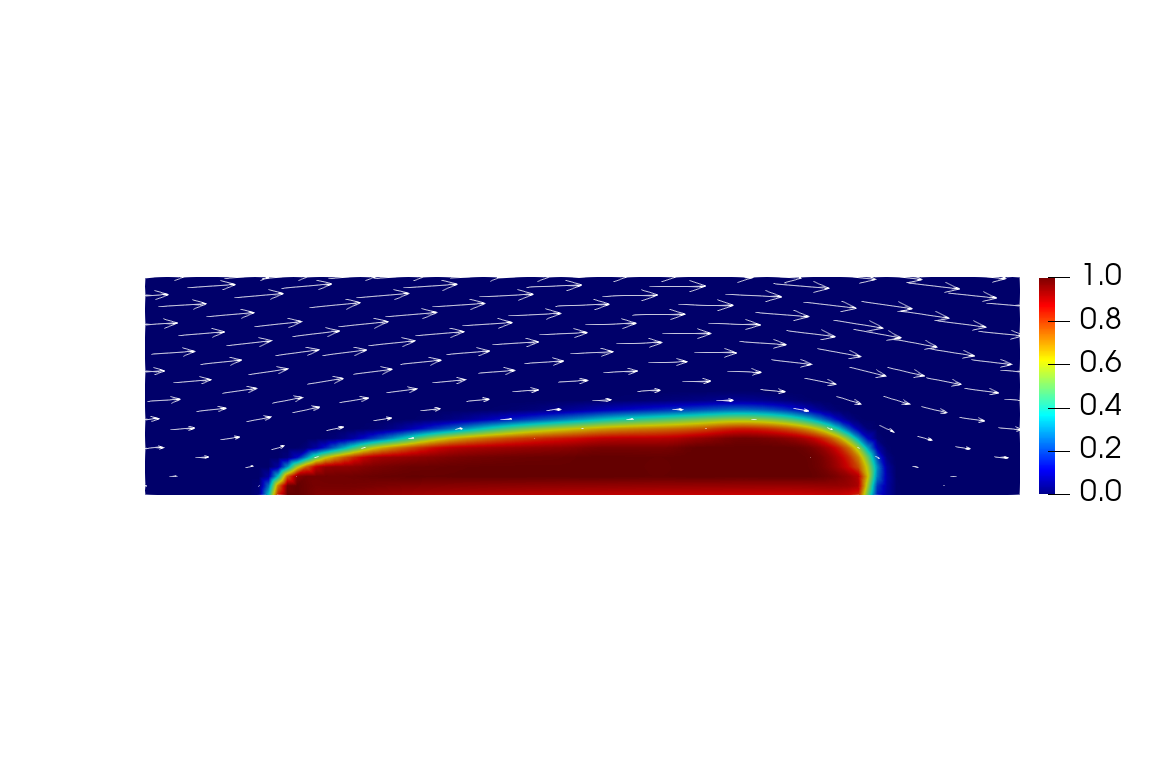}
\end{subfigure}
\begin{subfigure}[t]{0.38\textwidth}
\hspace*{0.35cm}  
\centering
    \includegraphics[width=\textwidth,trim={5cm 9cm 0cm 9cm},clip]{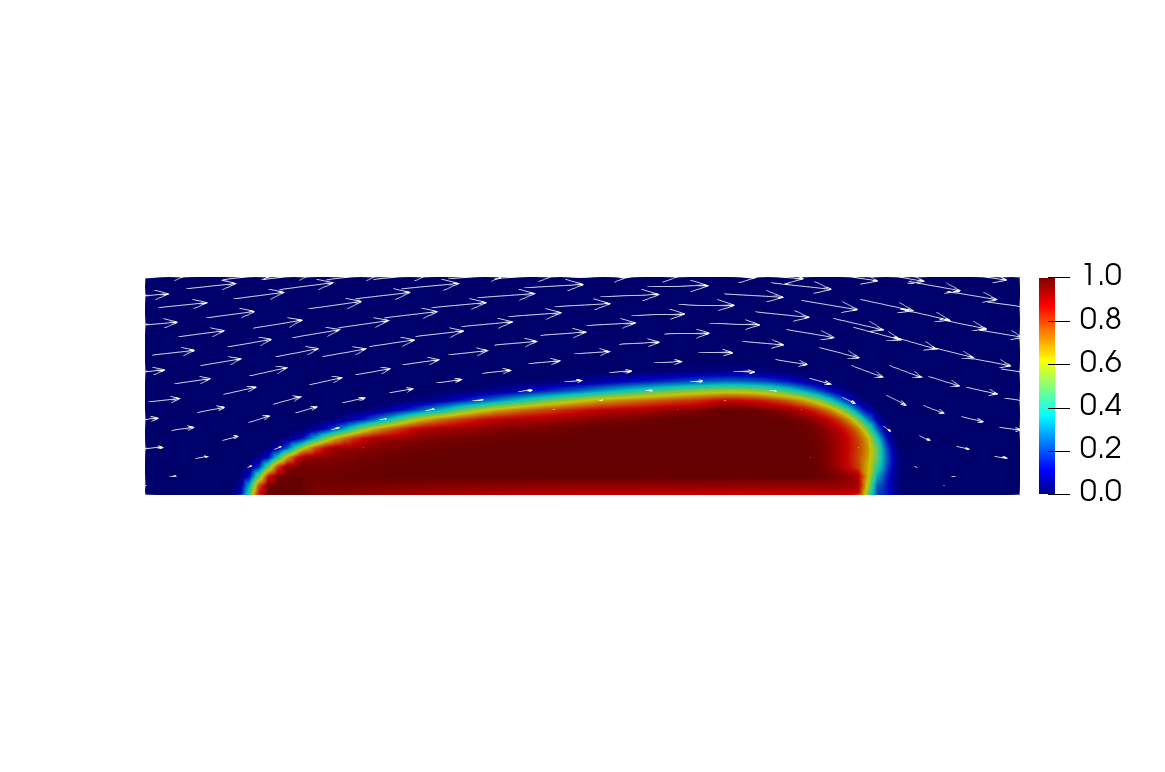}
\end{subfigure}
}

\caption{Close-up of 140 $\mu$m long by 35 $\mu$m high region around the thrombus to view various species (rows) at times 200, 400 and 600 seconds in the left, middle and right columns, respectively.  The first 3 rows show spatial concentrations (in nM) of $E_1$ and $S_2$ and the total thrombin concentration ($E_2 + E_2^b$), which was calculated in ParaView during post-processing.  The bottom row shows the resulting thrombus growth via the bound platelet fraction in a dynamic fluid environment where the fluid velocity field is depicted by the white arrows. }
\label{fig:exampleThrombosis}
\end{figure}

To verify the reliability of \emph{clotFoam} in modeling clotting phenomena, we compare its results with those in our previous work \cite{Leiderman2011grow,Leiderman2012} and simulate clotting in a 240 $\mu$m long by 60 $\mu$m high rectangular channel with an approximately 90  $\mu$m long adhesive and reactive patch centered on the bottom wall.  \edit{The domain is discretized using a uniform mesh of (128$\times$32) cells, which is divided into three blocks as depicted in Figure \ref{fig:injuryBlock_blockMesh} }. It should be noted that the coagulation reactions used in \emph{clotFoam} are a simplified version of those used in our previous work, and therefore, the outcomes are not expected to be identical. Nevertheless, we demonstrate that \emph{clotFoam} produces similar concentrations of bound thrombin, clot sizes and densities, all on the same timescale as in our previous studies \cite{Leiderman2011grow, Leiderman2012}.
This is illustrated in Figure \ref{fig:exampleThrombosis}, where we show spatial distributions of substrates, enzymes, and clots formed after 200, 400, and 600 seconds of clotting activity.

\subsection{Hemostasis in a microfluidic device} \label{sec:Illustrative-Example-Hemostasis}
\begin{figure}[H]

\makebox[\linewidth][c]{%
\begin{subfigure}[t]{0.50\textwidth}
\centering
    \includegraphics[width=\textwidth,trim={3cm 8cm 9cm 2cm},clip]{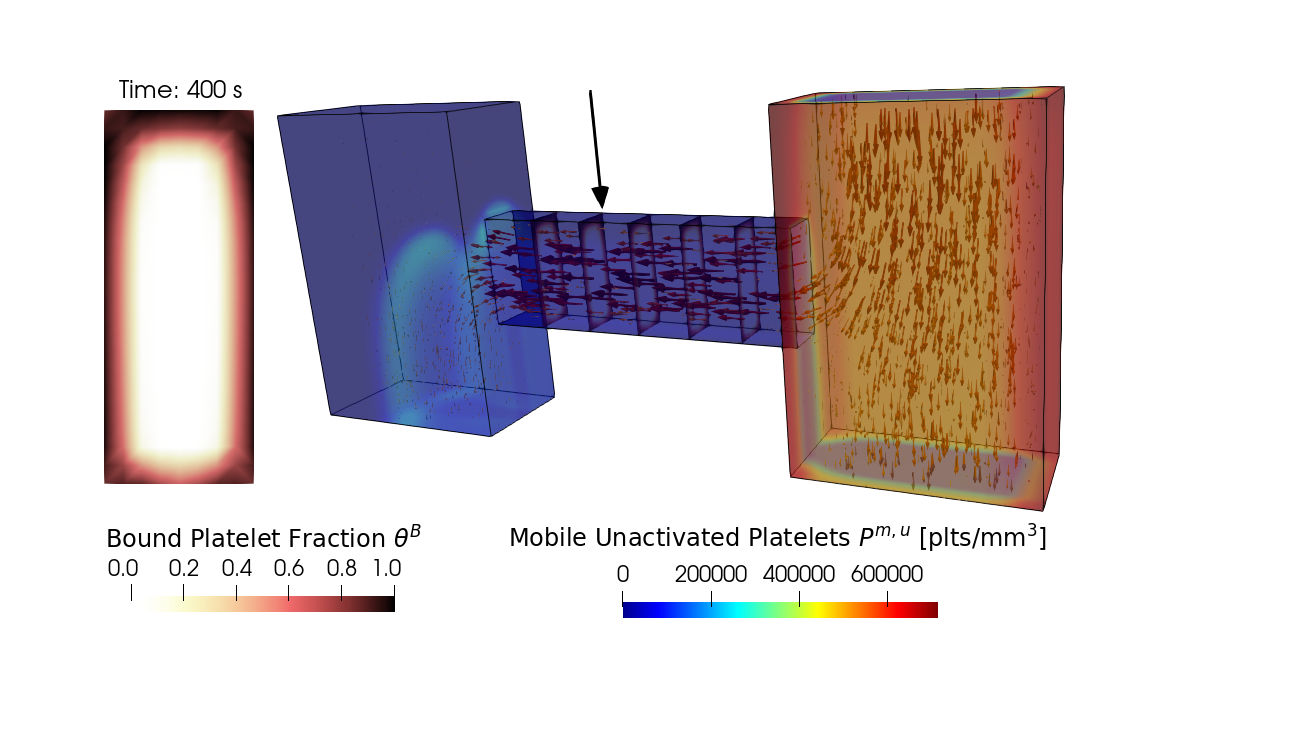}
\end{subfigure}
\begin{subfigure}[t]{0.50\textwidth}
\hspace*{0.35cm}  
\centering
    \includegraphics[width=\textwidth,trim={3cm 8cm 9cm 2cm},clip]{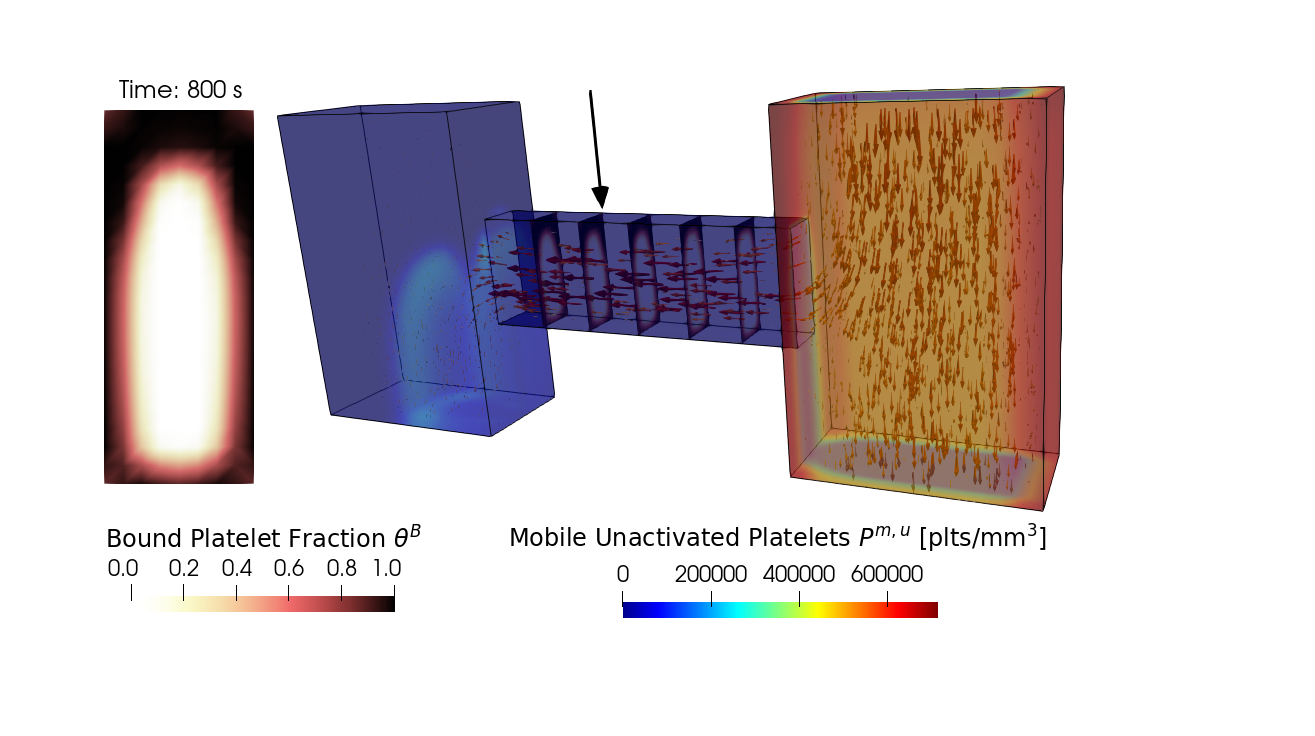}
\end{subfigure}
}

\makebox[\linewidth][c]{%
\begin{subfigure}[t]{0.50\textwidth}
\centering
    \includegraphics[width=\textwidth,trim={3cm 8cm 9cm 2cm},clip]{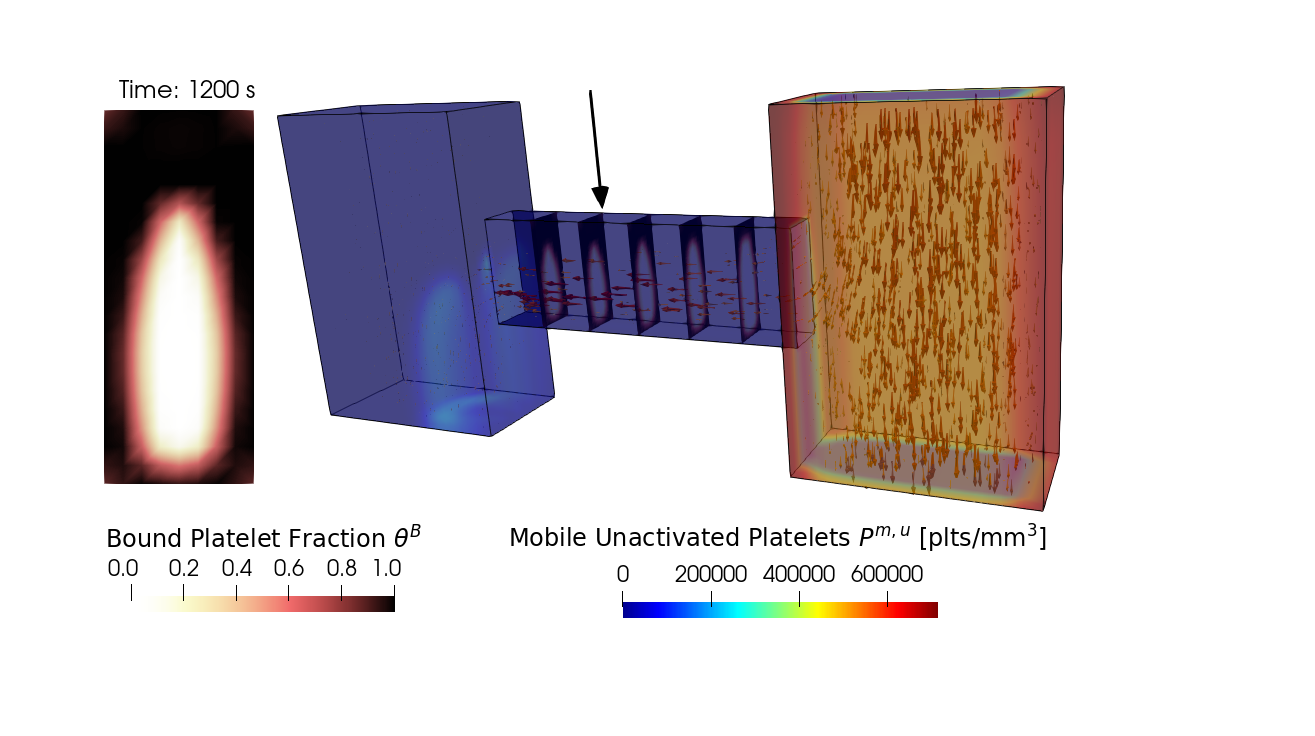}
\end{subfigure}
\begin{subfigure}[t]{0.50\textwidth}
\hspace*{0.35cm}  
\centering
    \includegraphics[width=\textwidth,trim={3cm 8cm 9cm 2cm},clip]{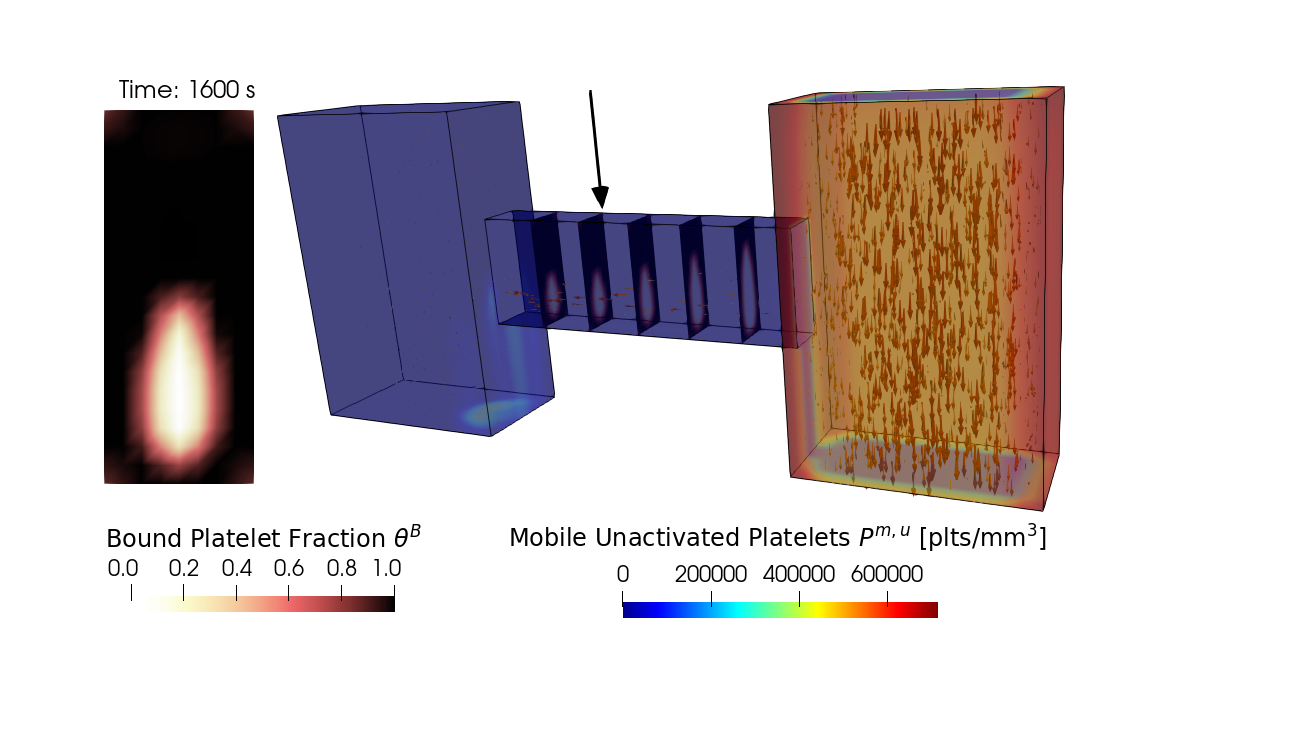}
\end{subfigure}
}
\makebox[\linewidth][c]{%
\begin{subfigure}[t]{0.6\textwidth}
\centering
    \includegraphics[width=\textwidth,trim={3cm 2cm 9cm 18.1cm},clip]{figs/speciesPlots/H_sideBySide_t_1200s.png}
\end{subfigure}
}
\caption{Snapshots of a \emph{clotFoam} simulation of hemostasis in an H-shaped microfluidic device for time 400, 800, 1200, 1600 seconds. The simulation shows mobile platelets entering the right vertical channel and being transported by the fluid (indicated by arrows) through the horizontal injury channel. As they pass through the injury, they start to aggregate and form a platelet plug, which is visualized via the bound platelet fraction $\theta^B$ in the five slices within the injury channel. }
\label{fig:exampleHemostasis}
\end{figure}

To demonstrate the versatility of \emph{clotFoam} in simulating clotting in various two and three dimentional domains, we have included a case that replicates the H-shaped microfluidic device used in Schoeman et al. \cite{Schoeman2016} to model hemostasis. In their experiments, whole blood is introduced into the right "blood channel," while a buffer fluid is introduced into the left vertical channel. The pressure difference between the channels causes blood to flow through the horizontal "injury channel," which is coated with tissue factor and collagen proteins that initialize coagulation and platelet adhesion/aggregation respectively. Clots build up in the injury channel without restricting blood flow in the blood channel. Figure \ref{fig:exampleHemostasis} shows snapshots of the 3D domain at four different times during the clotting process. Due to difficulties visualing the entire clot in the injury channel, we have displayed slices through the injury channel to highlight the spatial growth of the clot at various points down the injury channel. The rectangular plots to the left of each 3D domain are an enlarged view of the slice at the location indicated by the black arrow and show the clot distribution and its dynamics over time.
The thrombus growth patterns observed in the experimental results presented by Schoeman et al. \cite{Schoeman2016} show thrombus formation primarily at the front (right) of the injury channel while the thrombus generated by \emph{clotFoam}, shown in Figure \ref{fig:exampleHemostasis}, shows the buildup of thrombus closer to the back (left) of the injury channel. We hypothesize that this discrepancy may be attributed to the absence of shear dependence in the current platelet aggregation model, and this aspect will be investigated as part of our future work.  

\section{Impact and conclusions}
\label{sec:Impact-conclusions}
Computational models that simulate clotting phenomena have provided a significant step toward a better understanding of hemostasis, thrombosis, and clotting disorders. In this work, we have presented \emph{clotFoam}, an open-source software for simulating clotting using the computational fluid dynamics framework OpenFOAM. To demonstrate the reliability of \emph{clotFoam}, we compared its outcomes with our previous computational work studying thrombosis in rectangular channels \cite{Leiderman2011grow,Leiderman2012,Schoeman2016,Danes}. Our results show that \emph{clotFoam} produces a similar  clot structure and thrombin concentrations, thereby verifying its implementation of a reduced coagulation and platelet aggregation model. Furthermore, we demonstrated the versatility of \emph{clotFoam} by simulating clotting in an H-shaped microfluidic device, which was used in experiments for modeling hemostasis. While the simulations did not replicate the experiments exactly, they illustrated the potential of \emph{clotFoam} for investigating clotting in various microfluidic geometries.

In conclusion, \emph{clotFoam} offers a reliable and flexible platform for simulating clotting phenomena. With the ability to manipulate the platelet and coagulation cascade models, researchers can use \emph{clotFoam} to investigate various aspects of thrombus formation and design microfluidic devices for studying hemostasis. The open-source nature of \emph{clotFoam} also allows for community-driven development and improvement of the software, making it an accessible tool for researchers in the field.

\section*{Conflict of Interest}
No conflict of interest exists:
We wish to confirm that there are no known conflicts of interest associated with this publication and there has been no significant financial support for this work that could have influenced its outcome.

\section*{Acknowledgements}
\label{sec:acknowledgements}
We would like to thank our collaborators for their support and guidance: Aaron Fogelson, Suzanne Sindi, Keith Neeves, Dougald Monroe, and Michael Stobb. This work was, in part, supported by the National Institutes of Health (R01 HL151984), and the National Science Foundation CAREER program (DMS-1848221).





\appendix

\section{Platelet Aggregation Model }
\label{appendix:plateletAggregationModel}

From our previous work, the number densities of the four platelet species are denoted by $P^{m,u}, \; P^{m,a}, \; P^{b,a}, \; P^{b,se}$, where the superscripts represent mobile unactivated, mobile activated, platelet-bound activated, and subendothelium-bound activated respectively. Platelet aggregation is described by the following system of ADR equations:  
\begin{align}
	\pdv{P^{m,u}}{t} &= - \underbrace{\div \{  W(\theta^T) ( \u P^{m,u} - D_P \grad P^{m,u} )  \}}_\text{Transport by advection and "diffusion"}  \label{eq:Pmu_eqtn}\\
	&\;\;\; \; - \underbrace{\kadh H_\text{adh}(\x)( \Pmax - P^{b,se} ) P^{m,u}}_\text{Adhesion to subendothelium} 
                   - \underbrace{\{ A_1([\ADP]) + A_2(E_2) \}P^{m,u}}_\text{Activation by ADP or thrombin}, \nonumber \\[15pt]
	\pdv{P^{m,a}}{t} &= - \div \{ W(\theta^T) ( \u P^{m,a} - D_P \grad P^{m,a} ) \} - \kadh(\x)( \Pmax - P^{b,se} ) P^{m,a} \label{eq:Pma_eqtn}\\
	&\;\;\;\;    + \{ A_1([\ADP]) + A_2(E_2) \}P^{m,u} - \underbrace{\kcoh g(\eta)\Pmax P^{m,a}}_\text{Cohesion to bound platelets} \nonumber, \nonumber \\[15pt]
	\pdv{P^{b,a}}{t} &= -\kadh H_\text{adh}(\x)( \Pmax - P^{b,se}) P^{b,a} + \kcoh g(\eta)\Pmax P^{m,a}, \\[15pt]
	\pdv{P^{b,se}}{t} &= \kadh H_\text{adh}(\x)( \Pmax - P^{b,se}) (P^{m,u} + P^{m,a} + P^{b,a}) \label{eq:Pbse_eqtn}.
\end{align}
The hindered transport function is a monotonically decreasing function defined as $W(\theta^T) = \tanh(\pi(1-\theta^T))$, where $\theta^T$ is the ratio of the sum of all platelet species to $\Pmax$. The function assumes that the transport of mobile platelets is only slightly hindered until $\theta^T$ approaches approximately 0.5, after which the transport of mobile platelets is drastically reduced. The adhesion function $H_\text{adh}(\x)$ is chosen to be unity or zero, and defines the region where mobile platelets can stick to specified walls on the domain boundary.

Platelet-platelet cohesion is modeled through the parameters $\kcoh$ and the binding-affinity function $g(\eta)$, where the latter depends on a non-dimensional virtual substance $\eta$ that is produced by bound platelets.  In \emph{clotFoam}, the virtual substance is modeled by diffusing the bound platelet fraction a distance of $L_\eta $ over an interval of $\Delta t $ through the use of a discrete diffusion equation:
\begin{align}
\frac{\eta - \eta_0}{ \Delta t} = D_\eta \nabla^2 \eta, \label{eq:eta}
\end{align}
where $\eta_0$ is the current known bound platelet fraction, $\eta_0 = \theta^B_n$.  The diffusion constant is defined as $D_\eta = \frac{L_\eta^2}{4 \Delta t}$, and therefore \eqref{eq:eta} is implemented as:
\begin{align}
\eta - \frac{{L_\eta }^2}{4} \nabla^2{\eta} = \theta^B_n.
\end{align}
The definition of $\eta$ presented above departs somewhat from the corresponding quantity in our previous work, and is intended to rectify the overly diffusive approach used in our previous work \cite{Leiderman2011grow,Leiderman2012,Danes}. The binding affinity function is defined as:
\begin{equation}
g(\eta) =   \frac{g_0 \, (\eta - \eta_t)^3}{ \eta_\ast^3 + (\eta - \eta_t)^3} ,
\end{equation}
where $\eta_t$ is a threshold value for which there is no binding, $\eta_\ast + \eta_t$ indicates the value of $\eta$ for which $g(\eta)$ changes rapidly, and $g_0 = \frac{\eta_\ast^3 + (1 - \eta_t)^3}{(1 - \eta_t)^3} $ so that $g(1) = 1$.

\section{Coagulation model}
\label{appendix:coagulationModel}
The reduced coagulation network is modeled by the following set of enzymatic reactions: 
\begin{align}
S_1 + E_0 \xrightleftharpoons[k_{C_0}^-]{k_{C_0}^+} & \;C_0 \xrightarrow{k_{C_0}^\text{cat}} E_0 + E_1 \label{eq:reactionFirst}\\
S_1 + P_{1} &\xrightleftharpoons[k_{S_1}^-]{k_{S_1}^+} S_1^b \\
E_1 + P_{1} &\xrightleftharpoons[k_{E_1}^-]{k_{E_1}^+} E_1^b \\
S_2 + P_{2} &\xrightleftharpoons[k_{S_2}^-]{k_{S_2}^+} S_2^b \\
E_2 + P_{2} &\xrightleftharpoons[k_{E_2}^-]{k_{E_2}^+} E_2^b  \\
S_2^b + E_1^b \xrightleftharpoons[k_{C_1}^-]{k_{C_1}^+} &\;C_1 \xrightarrow{k_{C_1}^\text{cat}} E_1^b + E_2^b \label{eq:reactionFeedback1}\\
S_1^b + E_2^b \xrightleftharpoons[k_{C_2}^-]{k_{C_2}^+} &\;C_2 \xrightarrow{k_{C_2}^\text{cat}} E_1^b + E_2^b 
\label{eq:reactionFeedback2}
\end{align}
Using the law of mass action, a system of twelve differential equations is derived to govern the reactions of the twelve biochemical species present in the coagulation cascade. These equations are categorized as fluid-phase, platelet-bound, and subendothelium-bound for organizational purposes. All concentrations of the biochemical species are measured in units of nmol/mm$^3$, and platelet densities are converted to concentrations using Avogadro's constant $N_A = 6.02214076 \times 10^{14}$ nmol$^{-1}$, where $N_A$ is the number of molecules in one mole of a substance.

\subsection*{Fluid-phase:} 
\begin{align}
\pdv{S_1}{t} &= - \div (\u S_1 - D \grad S_1)  \\
&\;\;\;\;  - k_{S_1}^+ \Big\{\tfrac{N_{1}}{N_A} (P^{b,a} +  P^{se,a})-(S_1^b + E_1^b + C_1 + C_2) \Big\} S_1 + k_{S_1}^-S_1^b, \nonumber \label{fluidPhase_eqtn1}\\
& \hspace{5mm}\text{ with }  -D\, \n \cdot \grad S_1 \big|_{\partial \Omega_\text{inj}} = -k_{C_0}^+ S_1 E_0 + k_{C_0}^- C_0, \nonumber \\
\pdv{S_2}{t} &= - \div (\u S_2 - D \grad S_2) \\
&\;\;\;\; - k_{S_2}^+ \Big\{\tfrac{N_{2}}{N_A} (P^{b,a} +  P^{se,a})-(S_2^b + E_2^b + C_1 + C_2) \Big\} S_2
+ k_{S_2}^-  S_2^b, \nonumber \\
\pdv{E_1}{t} &= - \div (\u E_1 - D \grad E_1)  \\
&\;\;\;\; - k_{E_1}^+\Big\{ \tfrac{N_{1}}{N_A} (P^{b,a} + P^{se,a}) - (S_1^b + E_1^b + C_1 + C_2)\Big\}E_1
 + k_{E_1}^-E_1^b, \nonumber \label{fluidPhase_eqtn3}\\
    & \hspace{5mm}\text{ with }  -D\,  \n \cdot \grad E_1  \big|_{\partial \Omega_\text{inj}} = k_{C_0}^\text{cat}C_0, \nonumber \\
\pdv{E_2}{t} &= - \div (\u E_2 - D \grad E_2)   \\
&\;\;\;\;-  k_{E_2}^+ \Big\{\tfrac{N_{2}}{N_A} (P^{b,a} +  P^{se,a})-(S_2^b + E_2^b +C_1 + C_2) \Big\} E_2 + k_{E_2}^- E_2^b.  \nonumber\label{fluidPhase_eqtn4}
\end{align}

\subsection*{Platelet-bound:} 
\begin{align}
\pdv{S_1^b}{t} &=  \underbrace{k_{S_1}^+ \Big\{\tfrac{N_{1}}{N_A} (P^{b,a} +  P^{se,a}) -(S_1^b + E_1^b + C_1 + C_2) \Big\} S_1 - k_{S_1}^- S_1^b}_\text{Binding and unbinding on platelet surface}
	\\
	& \;\;\;\;  -k_{C_2}^+ S_1^b E_2^b + k_{C_2}^- C_2, \nonumber \\
\pdv{S_2^b}{t} &=  k_{S_2}^+ \Big\{\underbrace{\tfrac{N_{2}}{N_A} (P^{b,a} +  P^{se,a})}_{\substack{\text{Total concentration} \\ \text{of binding sites}}} -\underbrace{(S_2^b + E_2^b + C_1 + C_2)}_{\substack{\text{Occupied} \\ \text{binding sites}}} \Big\} S_2 - k_{S_2}^- S_2^b
	\\
	& \;\;\;\;  -k_{C_1}^+ S_2^b E_1^b + k_{C_1}^- C_1, \nonumber \\
\pdv{E_1^b}{t} &= k_{E_1}^+\Big\{ \tfrac{N_{1}}{N_A} (P^{b,a} + P^{se,a}) - (S_1^b + E_1^b + C_1 + C_2)\Big\}E_1  
- k_{E_1}^-E_1^b\\ 
& \;\;\;\;  -k_{C_1}^+ S_2^b E_1^b  + (k_{C_1}^- + k_{C_1}^\text{cat})C_1
+\underbrace{ k_{C_2}^\text{cat} C_2}_{\substack{\text{Positive} \\ \text{feedback}}}, \nonumber \\
\pdv{E_2^b}{t} &= k_{E_2}^+\Big\{ \tfrac{N_{2}}{N_A} (P^{b,a} + P^{se,a}) - (S_2^b + E_2^b + C_1 + C_2)\Big\}E_2  
- k_{E_2}^-E_2^b\\ 
& \;\;\;\;  -k_{C_2}^+ S_1^b E_2^b  + (k_{C_2}^- + k_{C_2}^\text{cat})C_2
+ \underbrace{k_{C_1}^\text{cat} C_1}_{\substack{\text{Positive} \\ \text{feedback}}}, \nonumber \\
\pdv{C_1}{t} &= k_{C_1}^+ S_2^b E_1^b - (k_{C_1}^- + k_{C_1}^\text{cat})C_1, \\
\pdv{C_2}{t} &= k_{C_2}^+ S_1^b E_2^b - (k_{C_2}^- + k_{C_2}^\text{cat})C_2.
\end{align}

\subsection*{Subendothelium-bound:}  
\begin{align}
\pdv{E_0}{t} &= -k_{C_0}^+ S_1 E_0 + (k_{C_0}^- + k_{C_0}^\text{cat})C_0, \text{ on } \partial \Omega_\text{inj}, \label{eq:seBound_E0}\\
\pdv{C_0}{t} &= k_{C_0}^+ S_1 E_0 - (k_{C_0}^- + k_{C_0}^\text{cat})C_0, \text{ on } \partial \Omega_\text{inj}.
\label{eq:seBound_C0}
\end{align}

\edit{ 
\section{Boundary conditions} \label{appendix:boundary-conditions}
The boundary conditions for most variables remain consistent across different geometries. The variables that experience changes in their boundary conditions when the geometry is altered are those introduced at the domain inlet, such as the fluid velocity $\u$, the mobile unactivated platelets $P^{m,u}$, and the kinematic pressure, $\pt = \p / \rho$, at the outlet. The specific conditions for these variables are detailed in the corresponding subsections specific to each domain. For all other variables, the following boundary conditions are applied:
\begin{table}[H]
\centering
\caption{Boundary conditions for each variable in the simulation. The table specifies the mathematical boundary condition and the corresponding implementation in OpenFOAM for each boundary. Some variables with the same boundary conditions have been grouped together for brevity. } \label{tab:generalBCs}
	\scriptsize
	\def\arraystretch{1.5}
	\begin{tabular}{|c c c c c|}
	\hline
        Variable(s) & \texttt{inlet} & \texttt{outlet} & \texttt{fixedWalls} & \texttt{injuryWalls} \\
        \hline
        $\u$ 
        & 
        \textit{see below}
        & 
        \begin{tabular}{c}
             $\pdv{\u}{\n} = 0 $ \\
             \texttt{zeroGradient}
        \end{tabular} 
        & 
        \begin{tabular}{c}
             $\u = 0$  \\
             \texttt{noSlip}
        \end{tabular} 
         & 
         \begin{tabular}{c}
             $\u = 0$  \\
             \texttt{noSlip}
        \end{tabular} 
        \\
        \hline
        $\pt$ 
        & 
        \begin{tabular}{c}
             $\pdv{\pt}{\n} = 0 $ \\
             \texttt{zeroGradient}
        \end{tabular} 
        & 
        \begin{tabular}{c}
             $\pt = \p_\text{out}$  \\
             \texttt{fixedValue}
        \end{tabular} 
        & 
        \begin{tabular}{c}
             $\pdv{\pt}{\n} = 0 $ \\
             \texttt{zeroGradient}
        \end{tabular} 
         & 
         \begin{tabular}{c}
             $\pdv{\pt}{\n} = 0 $ \\
             \texttt{zeroGradient}
        \end{tabular}  
        \\
        \hline
        $P^{m,u}$ 
        & 
        \begin{tabular}{c}
              \textit{see below} \\
             \texttt{codedFixedValue}
        \end{tabular} 
        & 
        \begin{tabular}{c}
             $\pdv{P^{m,u}}{\n} = 0 $ \\
             \texttt{zeroGradient}
        \end{tabular}  
        & 
        \begin{tabular}{c}
             $\pdv{P^{m,u}}{\n} = 0 $ \\
             \texttt{zeroGradient}
        \end{tabular} 
         & 
         \begin{tabular}{c}
             $\pdv{P^{m,u}}{\n} = 0 $ \\
             \texttt{zeroGradient}
        \end{tabular}
        \\
        \hline
        \begin{tabular}{c}
             $P^{m,a}$   \\
             $P^{b,a}$   \\
             $P^{b,se}$
        \end{tabular}
        & 
        \begin{tabular}{c}
             $\pdv{P^{k}}{\n} = 0 $ \\
             \texttt{zeroGradient}
        \end{tabular} 
        & 
        \begin{tabular}{c}
             $\pdv{P^{k}}{\n} = 0 $ \\
             \texttt{zeroGradient}
        \end{tabular}  
        & 
        \begin{tabular}{c}
             $\pdv{P^{k}}{\n} = 0 $ \\
             \texttt{zeroGradient}
        \end{tabular}  
         & 
         \begin{tabular}{c}
             $\pdv{P^{k}}{\n} = 0 $ \\
             \texttt{zeroGradient}
        \end{tabular} 
        \\
        \hline
        $S_1$
        & 
        \begin{tabular}{c}
             $S_1 = \binom{\text{normal}}{\text{concentration}}$  \\
             \texttt{fixedValue}
        \end{tabular} 
        & 
        \begin{tabular}{c}
             $\pdv{S_1}{\n} = 0 $ \\
             \texttt{zeroGradient}
        \end{tabular}  
        & 
        \begin{tabular}{c}
             $\pdv{S_1}{\n} = 0 $ \\
             \texttt{zeroGradient}
        \end{tabular}  
         & 
         \begin{tabular}{c}
             $-D\pdv{S_1}{\n} = -k_{C_0}^+S_1 + k_{C_0}^-C_0 $ \\
             \texttt{codedMixed} 
        \end{tabular} 
        \\
        \hline
        $E_1$
        & 
        \begin{tabular}{c}
             $\pdv{E_1}{\n} = 0 $ \\
             \texttt{zeroGradient}
        \end{tabular} 
        & 
        \begin{tabular}{c}
             $\pdv{E_1}{\n} = 0 $ \\
             \texttt{zeroGradient}
        \end{tabular}  
        & 
        \begin{tabular}{c}
             $\pdv{E_1}{\n} = 0 $ \\
             \texttt{zeroGradient}
        \end{tabular}  
         & 
         \begin{tabular}{c}
             $-D\pdv{E_1}{\n} = k_{C_0}^\text{cat}C_0 $\\
             \texttt{codedMixed} 
        \end{tabular} 
        \\
        \hline
        $S_2$
        & 
        \begin{tabular}{c}
             $S_2 = \binom{\text{normal}}{\text{concentration}}$  \\
             \texttt{fixedValue}
        \end{tabular} 
        & 
        \begin{tabular}{c}
             $\pdv{S_2}{\n} = 0 $ \\
             \texttt{zeroGradient}
        \end{tabular}  
        & 
        \begin{tabular}{c}
             $\pdv{S_2}{\n} = 0 $ \\
             \texttt{zeroGradient}
        \end{tabular}  
         & 
         \begin{tabular}{c}
             $\pdv{S_2}{\n} = 0 $ \\
             \texttt{zeroGradient}
        \end{tabular}  
        \\
        \hline
        $E_2$
        & 
        \begin{tabular}{c}
             $\pdv{E_2}{\n} = 0 $ \\
             \texttt{zeroGradient}
        \end{tabular}   
        & 
        \begin{tabular}{c}
             $\pdv{E_2}{\n} = 0 $ \\
             \texttt{zeroGradient}
        \end{tabular}     
        & 
        \begin{tabular}{c}
             $\pdv{E_2}{\n} = 0 $ \\
             \texttt{zeroGradient}
        \end{tabular}     
         & 
         \begin{tabular}{c}
             $\pdv{E_2}{\n} = 0 $ \\
             \texttt{zeroGradient}
        \end{tabular}     
        \\
        \hline
        \begin{tabular}{c}
             $S_1^b$   \\
             $S_2^b$  \\
             $E_1^b$  \\
             $E_2^b$  \\
             $C_1$  \\
             $C_2$  
        \end{tabular}
        & 
        \begin{tabular}{c}
             $\pdv{C^k}{\n} = 0 $ \\
             \texttt{zeroGradient}
        \end{tabular}
        & 
        \begin{tabular}{c}
             $\pdv{C^k}{\n} = 0 $ \\
             \texttt{zeroGradient}
        \end{tabular}  
        & 
        \begin{tabular}{c}
             $\pdv{C^k}{\n} = 0 $ \\
             \texttt{zeroGradient}
        \end{tabular} 
         & 
         \begin{tabular}{c}
             $\pdv{C^k}{\n} = 0 $ \\
             \texttt{zeroGradient}
        \end{tabular}  
        \\
        \hline
        \begin{tabular}{c}
             $E_0$  \\
             $C_0$
        \end{tabular}
        & 
        \begin{tabular}{c}
             $\pdv{C^k}{\n} = 0 $ \\
             \texttt{zeroGradient}
        \end{tabular}   
        & 
        \begin{tabular}{c}
             $\pdv{C^k}{\n} = 0 $ \\
             \texttt{zeroGradient}
        \end{tabular}     
        & 
        \begin{tabular}{c}
             $\pdv{C^k}{\n} = 0 $ \\
             \texttt{zeroGradient}
        \end{tabular}     
         & 
         \begin{tabular}{c}
             See equations \eqref{eq:seBound_E0} and \eqref{eq:seBound_C0} \\
             solved in \texttt{odeSolver.H}
        \end{tabular}     
        \\
        \hline
        ADP
        & 
        \begin{tabular}{c}
             $\pdv{[\ADP]}{\n} = 0 $ \\
             \texttt{zeroGradient}
        \end{tabular}   
        & 
        \begin{tabular}{c}
             $\pdv{[\ADP]}{\n} = 0 $ \\
             \texttt{zeroGradient}
        \end{tabular}     
        & 
        \begin{tabular}{c}
             $\pdv{[\ADP]}{\n} = 0 $ \\
             \texttt{zeroGradient}
        \end{tabular}     
         & 
         \begin{tabular}{c}
             $\pdv{[\ADP]}{\n} = 0 $ \\
             \texttt{zeroGradient}
        \end{tabular}     
        \\
        \hline
 \end{tabular}
\end{table}

\subsection{Boundary conditions for thrombosis in a rectangular channel} \label{appendix:2DrectangleBC}
The velocity profile for $\u = [u\; v]^T$ is prescribed using the following function, with a shear rate  of $\shear = 1000$ s$^{-1}$ and radius $r = 30 \;\mu$m:
\begin{equation}
    u(y) =  -\frac{\shear}{2r}(y - r)^2 + \frac{1}{2} \shear r  \label{eq:parabolicVelocity}.
\end{equation}
This parabolic profile can be implemented in OpenFOAM using the \texttt{codedFixedValue} boundary condition

The mobile-unactivated platelets $P^{m,u}$ exhibit margination behavior at the inlet, where the concentration near the wall is higher compared to the center of the vessel. To capture this phenomenon, we employ a shape function based on the inlet profile proposed by Eckstein and Belgacem \cite{eckstein1991model}.  The inlet profile for mobile-unactivated platelets entering a 2D computational domain $\Omega = [x_0,\;x_\text{max}] \times [y_0,\;y_\text{max}]$ is described by:
\begin{align}
P^{m,u}(x_0,y,t) = P_0 \, c(y), \label{eq:plateletShapeFunction2D}
\end{align}
where $P_0$ is the normal density of platelets. The shape function $c(y)$ is defined as:
\begin{align}
c(y) = C_0 \Big[ 1 + KR(y,r)^{m-1} (1 - R(y,r))^{n-1} \Big]. 
\end{align}
with $C_0$ being a normalizing parameter, $K$ determining the relative amplitude of the shape, and $m,\;n\in \mathbb{N}$. The function $R(y,r)$ is given by:
\begin{align}
R(y,r) = \frac{ | y - r | }{r},
\end{align}
where the vessel radius is $r = (y_\text{max}-y_0)/2$. The exponents are set as $m=19$ and $n=2$ to produce the effect of a near-wall peak added to a uniform density $P_0$. The normalization parameter is defined as:
\begin{align}
\frac{1}{C_0} = \frac{1}{2r}  \int_{y_0}^{y_\text{max}} \Big[ 1 + KR(y,r)^{m-1} (1 - R(y,r))^{n-1} \Big] \, dy 
\end{align}
Figure \ref{fig:plateletProfileP2} shows the specific inlet profile used in the thrombosis example in Section \ref{sec:Illustrative-Example-Thrombosis}, where $r = 30 \; \mu$m and $K = 330$.  

\begin{figure}[ht!]
\makebox[\linewidth][c]{%
  \includegraphics[width=0.75\textwidth,trim={0cm 0cm 0cm 0cm},clip]{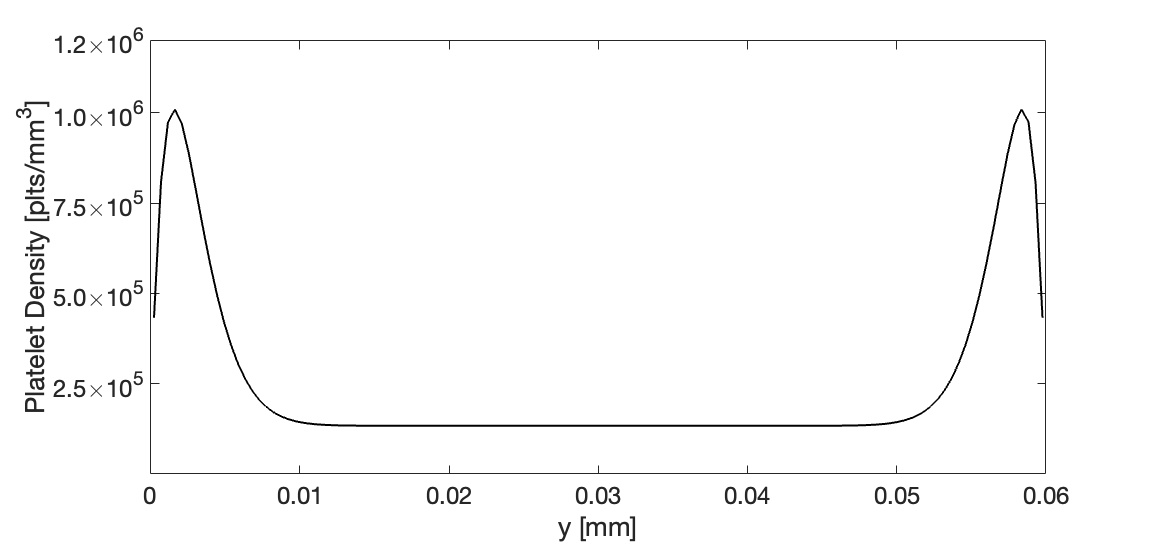}
}
\caption{Inlet profile for the mobile-unactivated platelets in a vessel with $r = 30 \; \mu$m.  The peak-to-centre ratios is set to approximately 7.5 using an amplitude parameter of $K = $  330. }
\label{fig:plateletProfileP2}
\end{figure}

\subsection{Boundary conditions for hemostasis in a microfluidic device} \label{appendix:3DrectangleBC}
The velocity profile $\u = [u\; v\; w]^T$ is prescribed using the \texttt{flowRateInletVelocity} boundary condition, with volumetric flow rates $Q_\text{left}$ and $Q_\text{right}$ at the upper left and upper right inlets, respectively. The kinematic pressure at the lower outlets is denoted by $\pt_\text{left}$ and $\pt_\text{right}$. The specific values for the flow rates and kinematic pressure, adapted from Schoeman et al. \cite{Schoeman2016}, are provided in the table below:

\begin{table}[H]
\centering
\caption{Fluid velocity and kinematic pressure boundary conditions.} \label{tab:BCs-3DHemostasis}
	\footnotesize
	\def\arraystretch{1.5}
	\begin{tabular}{|l c |}
	\hline
        Left inlet  ($Q_\text{left}$) & $1.17 \times 10^{-2} \; \text{mm}^3/\text{s}$\\
        Right inlet  ($Q_\text{right}$) & $9.32 \times 10^{-2}\; \text{mm}^3/\text{s}$\\
        Left outlet  ($\pt_\text{left}$) & $1.11 \; \times 10^6 \; \text{mm}^2 /\text{s}^2 $\\
        Right outlet  ($\pt_\text{right}$) & $1.78 \; \times 10^6 \; \text{mm}^2 /\text{s}^2$\\
	   		    	    \hline
	\end{tabular}
\end{table}

For the mobile-unactivated platelets, the shape function in equation \eqref{eq:plateletShapeFunction2D} is adapted for a 3D channel, $\Omega = [x_0,\;x_\text{max}] \times [y_0,\;y_\text{max}] \times [z_0,\;z_\text{max}]$.  The normalization constant $C_0$ is then calculated in two dimensions.  The shape function $c(x,z)$ is defined by:
{\small
\begin{align}
c(x,z) = C_0 \Big[ 1 +  K_x R(x,r_x)^{m-1} (1 - R(x,r_x))^{n-1}
 +  K_z R(z,r_z)^{m-1} (1 - R(z,r_z))^{n-1} \Big]
\end{align}
}
In the 3D hemostasis example, the amplitude parameters are set to $K_x = K_z = 547$ to create a peak-to-center ratio of approximately 11.9.  The exponents are set to $m = 19$ and $n = 2$.

\section{Discretization Schemes} \label{appendix:discretization}
The discretization schemes employed in this work are summarized in Table \ref{tab:discretization-schemes}.
 To limit spurious oscillations caused by steep gradients at the thrombus edges, the van Leer flux limiter \cite{VANLEER1974361} is applied in the advective terms for platelets and biochemical species. In Section \ref{sec:numericalMethods}, we provided a detailed discussion on the choice of interpolation scheme for the hindered transport function $W(\theta^T)$, which plays a crucial role in the simulation. In the table below, \texttt{Theta\_Tfa} and \texttt{Theta\_Tfd} represent the face values of $\theta^T$ used for hindered advection and hindered diffusion, respectively. Both simulations employ a variable time step $\Delta t$ that is set to ensure the Courant number, \texttt{maxCo}, does not exceed 0.75. For more practical examples and further information, please refer to the tutorials directory within the \emph{clotFoam} repository on GitHub.
\begin{table}[H]
\centering
\caption{Discretization schemes for orthogonal meshes in \emph{clotFoam}.} \label{tab:discretization-schemes}
	\scriptsize
 \def\arraystretch{1.5}
	\begin{tabular}{|c | c|}
	\hline
         \texttt{ddtSchemes}  & 
         \begin{tabular}{p{35mm}p{50mm}}
           \texttt{default}   &  \texttt{CrankNicolson 0.9;}\\[-2pt]
            \texttt{ddt(phi)}   &  \texttt{CrankNicolson 0.9;}  
         \end{tabular}\\
         \hline
         \texttt{gradSchemes} & 
         \begin{tabular}{p{35mm}p{50mm}}
           \texttt{default}   &  \texttt{Gauss linear;}\\[-2pt]
            \texttt{grad(p)}   &  \texttt{Gauss linear;}  
         \end{tabular}\\ 
         \hline
         \texttt{divSchemes} & 
         \begin{tabular}{p{35mm}p{50mm}}
           \texttt{default}   &  \texttt{none;}\\[-2pt]
            \texttt{div(phi,U)}   &  \texttt{Gauss linear;}  \\[-2pt]
            \texttt{div(phiPlt,Plt)} & \texttt{Gauss limitedVanLeer 0.0 \$Pmax;}  \\[-2pt]
            \texttt{div(phi,chems)} & \texttt{Gauss vanLeer;} 
         \end{tabular}\\
         \hline
         \texttt{laplacianSchemes} & 
         \begin{tabular}{p{35mm}p{50mm}}
           \texttt{default}   &  \texttt{Gauss linear orthogonal;}
         \end{tabular}\\
         \hline
         \texttt{interpolationSchemes} & 
         \begin{tabular}{p{35mm}p{50mm}}
           \texttt{default}   &  \texttt{linear}\\[-2pt]
            \texttt{interpolate(Theta\_Tfa)}   &  \texttt{downwind phi;}  \\[-2pt]
            \texttt{interpolate(Theta\_Tfd)} & \texttt{localMax;} 
         \end{tabular}\\
         \hline
         \texttt{snGradSchemes} & 
         \begin{tabular}{p{35mm}p{50mm}}
           \texttt{default}   &  \texttt{orthogonal;}
         \end{tabular}\\
	   		    	    \hline
	\end{tabular}
\end{table}

}

\section{Parameter Values}
\label{appendix:paramVals}
All of the parameters are from various sources as described in Leiderman \& Fogelson \cite{Leiderman2011grow}. 

\begin{table}[H]
\centering
\caption{Fluid equation parameters.}
	\footnotesize
	\def\arraystretch{1.5}
	\begin{tabular}{|l c |}
	\hline
        Fluid density ($\rho$) & 1.0 $\times 10^{-3}$ g/mm$^3$\\
        Dynamic viscosity $(\mu)$ & 2.62507 $\times 10^{-3}$ g/mm/s \\
        Kinematic viscosity ($\nu$) & 2.62507 mm$^2$/s\\
        Carman-Kozeny constant ($C_{CK}$) & 1.0 $\times 10^6$ mm$^{-2}$\\
	   		    	    \hline
	\end{tabular}
\end{table}

\begin{table}[H]
\centering
\caption{Diffusion coefficients by species.}

	\footnotesize
	\def\arraystretch{1.5}
	\begin{tabular}{|l c |}
	\hline
	Platelets ($D_P$) & $2.5 \times 10^{-5} \; \text{mm}^2/\text{s}$\\
	ADP ($D_\text{ADP}$)& $5 \times 10^{-4} \; \text{mm}^2/\text{s}$\\
	All other chemical species ($D$)&  $5 \times 10^{-5} \; \text{mm}^2/\text{s}$\\ 
	   		    	    \hline
	\end{tabular}
\end{table}

\begin{table}[H]
\centering
\caption{Platelet equation parameters.}

	\footnotesize
	\def\arraystretch{1.5}
	\begin{tabular}{|l c|}
	\hline
		Platelet diameter $( P_\text{diam} )$     &   3.0 $ \times 10^{-3} \text{ mm}$ \\

		Maximum packing density $(P_\text{max})$ & 6.67 $ \times 10^{7}$ platelets$/  \text{mm}^3 $  \\

		Normal density of platelets $(P_0)$ & 2.5 $ \times 10^{5}$ platelets$/  \text{mm}^3 $  \\

		Adhesion rate $(k_\text{adh})$     &   3.3212 $ \times 10^{-8} \text{ mm}^3 / \text{s} $  \\

		Cohesion  rate $(k_\text{coh}\times P_\text{max})$     &   1.0 $\times 10^{4} \; \text{s} ^{-1}$  \\

		Threshold for binding affinity $(\eta_t)$     &   1.0 $\times 10^{-1} $  \\

		Rapid change in binding affinity  $(\eta^\ast + \eta_t)$     &   $0.5 - \eta_t$   \\

            Length of diffusion for the virtual substance $(L_\eta)$ & $2.0 \times P_\text{diam}$ \\
	
		Rate of activation by ADP $(k^\text{pla}_\text{adp})$     &   0.34 $\text{s} ^{-1}$  \\
		
			Rate of activation by thrombin $(k^\text{pla}_{e_2})$     &   0.5 $\text{s} ^{-1}$  \\
			
			Critical concentration of
						ADP $([\ADP]^\ast)$     &   2.0 $\times 10^{-3} \; \text{nmol of ADP per mm}^3$  \\
						
						Critical concentration of
									thrombin $(e_2^\ast)$     &   1.0 $\times 10^{-6} \; \text{nmol of $e_2$ per mm}^3$ \\
		Total ADP released $(\hat A)$     &   $2.0\times 10^{-8} $ nmol of ADP per platelet \\
Total time of ADP secretion ($\tau_F$) & 6 s \\
Step size for calculating ADP secretion ($\Delta \tau$) & 0.25 s \\
							\hline
	\end{tabular}
\end{table}

\begin{table}[H]
\centering
\caption{Normal concentrations and surface binding site numbers.}
	\footnotesize
    \def\arraystretch{1.5}
	\begin{tabular}{|l c c c|}
	\hline
	Chemical & Concentration & Concentration & Binding Sites \\[-4pt]
	               & (clotFOAM) &  & \\
	\hline
	  $E_0$ (TF:VIIa) & $1.5 \times 10^{-7} $ nmol/mm$^2$
	    & 15 fmol/cm$^2$ & 
	    \\
	    $E_1$ (Xa) & 
	    	    &  & $N_{1}$ = 2700
	    	    \\
	   	$E_2$ (Thrombin) & 
	   		    	    &  & $N_{2}$ = 2000
	   		    	    \\

	    $S_1$ (X) & $1.7 \times 10^{-4} $ nmol/mm$^3$
	    	    & 0.17 $\mu$M & $N_{1}$ = 2700
	    	    \\
	   	$S_2$ (II) &  $1.4 \times 10^{-3} $ nmol/mm$^3$
	   		    	    & 1.4 $\mu$M & $N_{2} = 2000$ 
	   		    	    \\
	   		    	    \hline
	\end{tabular}
\end{table}

\begin{table}[H]
\centering
\caption{Reactions on subendothelium and platelet surface.}
	\footnotesize
	\def\arraystretch{1.5}
	\begin{tabular}{|l c c c c c c|}
	\hline
	Reaction &  Complex & Product&  mm$^{3}$nmol$^{-1}$s$^{-1}$ & M$^{-1}$ s$^{-1}$ & s$^{-1}$ & s$^{-1}$\\ [-4pt]
	Activation (of-, by-) & & &(clotFOAM) & & &\\
	\hline
	  ($S_1$, $E_0$) & $C_0$ & $E_1$ & $k_{C_0}^+ = 8.95\times 10^3$ & $k_{C_0}^+ = 8.95 \times 10^6$ & $k_{C_0}^- = 1.0$
	  & $k_{C_0}^\text{cat} = 1.15$
	    \\
	  ($S_2^b$, $E_1^b$) & $C_1$ & $E_2^b$ & $k_{C_1}^+ = 1.03 \times 10^{5}$ & $k_{C_1}^+ = 1.03 \times 10^8$ & $k_{C_1}^- = 1.0$
	  & $k_{C_1}^\text{cat} = 30.0$
	    \\
	    ($S_1^b$, $E_2^b$) & $C_2$ & $E_1^b$ & $k_{C_2}^+ = 1.73 \times 10^{4}$ & $k_{C_2}^+ = 1.73 \times 10^7$ & $k_{C_2}^- = 1.0$
	    	  & $k_{C_2}^\text{cat} = 0.23$
	    	    \\
	    \hline
	\end{tabular}
\end{table}

\begin{table}[H]
\centering
\caption{Binding on platelet surface}
	\footnotesize
	\def\arraystretch{1.5}
	\begin{tabular}{|l c c c c l|}
	\hline
	Reaction  & Reactants & Product&  mm$^{3}$nmol$^{-1}$s$^{-1}$ & M$^{-1}$ s$^{-1}$ & \multicolumn{1}{c|}{s$^{-1}$}\\[-4pt]
	 & & & (clotFOAM) & & \\
	 \hline
	 $S_1$ (V) & $S_1$, $P_{1}$ &$S_1^b$ & $k_{S_1}^+ = 5.7 \times 10^4$ & $k_{S_1}^+ = 5.7 \times 10^7$ & $k_{S_1}^- = 0.17$
	 	    \\
	 $E_1$ (Xa) & $E_1$, $P_{1}$ &$E_1^b$ & $k_{E_1}^+ = 1.0 \times 10^4$ & $k_{E_1}^+ = 1.0 \times 10^7$ & $k_{E_1}^- = 2.5 \times 10^{-2}$
\\
	  $S_2$ (II) & $S_2$, $P_{2}$ &$S_2^b$ & $k_{S_2}^+ = 1.0 \times 10^4$ & $k_{S_2}^+ = 1.0 \times 10^7$ & $k_{S_2}^- = 5.9$
	    \\
	 $E_2$ (IIa) & $E_2$, $P_{2}$ &$E_2^b$ & $k_{E_2}^+ = 1.0 \times 10^4$ & $k_{E_2}^+ = 1.0 \times 10^7$ & $k_{E_2}^- = 5.9$
	    	 	    \\
	    \hline
	\end{tabular}
\end{table}

\pagebreak
\noindent\textbf{\Large Supporting Information}

\section*{Mesh generation for 3D hemostasis example}
\begin{figure}[H]
\makebox[\linewidth][c]{%
\begin{subfigure}[b]{.59\textwidth}
  \centering
  \includegraphics[width=\textwidth,trim={4cm 0cm 4cm 0cm},clip]{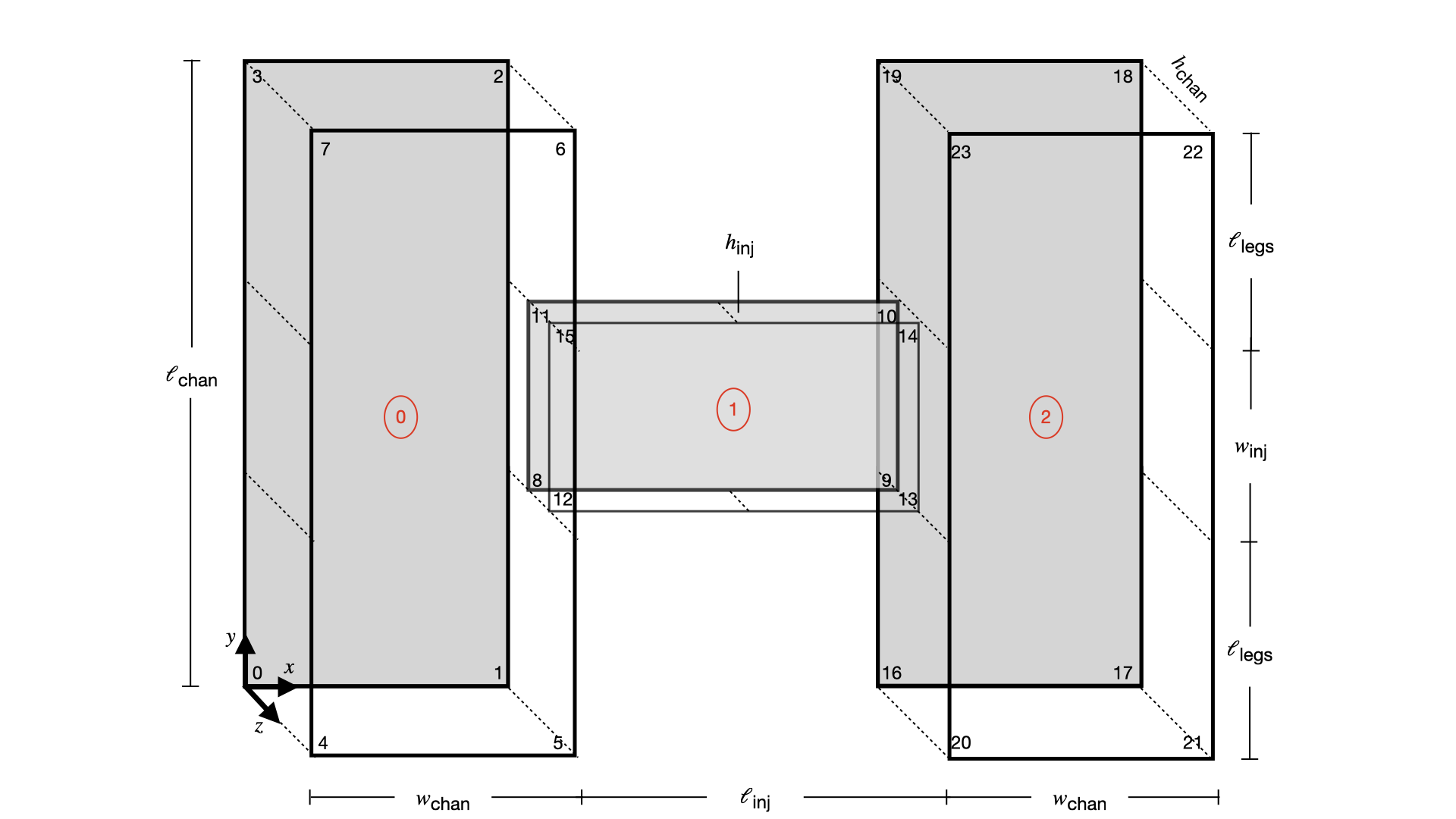}
\end{subfigure}%

\begin{subfigure}[b]{.59\textwidth}
  \centering
  \includegraphics[width=0.95\textwidth]{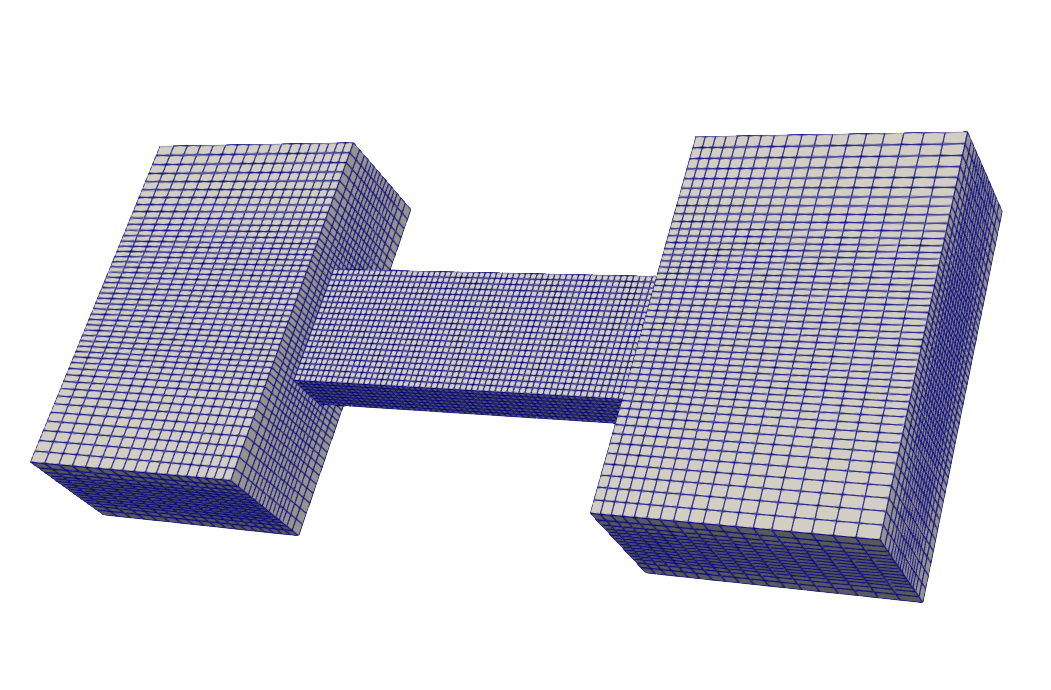}
\end{subfigure}%
}
\caption{Meshing the H-shaped microfluid device.} \label{fig:H-domain}
\end{figure}

To define the mesh in OpenFOAM, we utilize 3 blocks as seen in Figure \ref{fig:H-domain}. The creation of holes in the vertical channels at the interface with the injury channel is accomplished using the \texttt{mergePatchPairs} function. The dimensions of the H-domain, which serve as the computational domain, are provided in Table \ref{tab:H-domain-dimensions}, following the description given in Schoeman et al. \cite{Schoeman2016}.
The mesh is defined so that the cells in the injury channel have a uniform spacing $\Delta x = \Delta y = \Delta z = $2.5 $\mu$m.  The mesh outside of the injury channel is graded for computational efficiency.  The resulting mesh, as shown in Figure \ref{fig:H-domain}, consists of 37,248 finite volume cells.  The specific grading parameters used to generate the mesh can be found in the file \texttt{tutorials/Hjunction3D/system/blockMeshDict} in the \emph{clotFoam} repository on GitHub. These parameters were calculated using the free blockMesh grading tool \cite{OpenFOAM-wiki-grading}.

\begin{table}[H]
\centering
\caption{Dimensions of the computational domain of the H-shaped microfluidic device.} \label{tab:H-domain-dimensions}
	\small
	\def\arraystretch{1.5}
	\begin{tabular}{|l  c|}
	\hline
	Width of vertical channel ($w_\text{chan}$) & 100 $\mu$m \\
	Height of vertical channel ($h_\text{chan}$) & 50 $\mu$m \\
	Length of vertical channel ($\ell_\text{chan}$) & 150 $\mu$m \\
	Width of injury channel ($w_\text{inj}$) & 50 $\mu$m \\
	Height of injury channel ($h_\text{inj}$) & 20 $\mu$m \\
	Length of injury channel ($\ell_\text{inj}$) & 150 $\mu$m \\
	\hline
	\end{tabular}
\end{table}

\section*{Mesh convergence study}
To evaluate the accuracy and convergence of the \emph{clotFoam} solver, we performed a mesh convergence study in a rectangular domain measuring 120 $\mu$m in length, 30 $\mu$m in height, and with an injury length of 45 $\mu$m. The objective of this study was to assess the error by comparing the solution between a fine mesh and a coarse mesh. We utilized five different mesh discretizations, with uniform cell heights and widths ($h$) of 3, 1.5, 0.75, 0.375, and 0.1875 $\mu$m.  
In this study, the finest mesh was denoted as $h_\text{fine} = 0.1875 \; \mu$m, and its corresponding solution was represented by $\psi_\text{fine}$. 
To facilitate the comparison of solutions across different mesh resolutions, we employed the \texttt{Resample with Dataset} filter in ParaView to interpolate the coarse meshes onto the finest mesh. To ensure numerical stability, the time step was selected as a function of $h$ in a way that the maximum Courant number ($C_0$) did not exceed 0.75 for each mesh. 
The Courant number is defined as $C_0 = \frac{ \Delta t \|\u\| }{ h }$, where $\Delta t$ represents the time step and $\|\u\|$ is the magnitude of the velocity field.

The relative error for each solution $\psi_{h_k}$ with mesh $h_k$ is calculated using the following formula:
\begin{equation}
    E(\psi_{h_k}) = \frac{\| \psi_\text{fine} - \psi_{h_k} \|_{\ell_2}}{ \| \psi_\text{fine} \|_{\ell_2}},
\end{equation}
where the $\ell_2$ norm is defined for a solution vector $\x \in \mathbb{R}^n$  as:
\begin{align}
 \| \x \|_{\ell_2} &= \sqrt{ \sum_{i = 1}^n  x_i^2}, \label{eq:l2-norm} 
\end{align}
The convergence rate, denoted by $r$, was approximated by fitting a line to the error vectors on a log-log scale using least squares.  

\begin{figure}[H]
\makebox[\linewidth][c]{%
  \includegraphics[width=0.5\textwidth,]{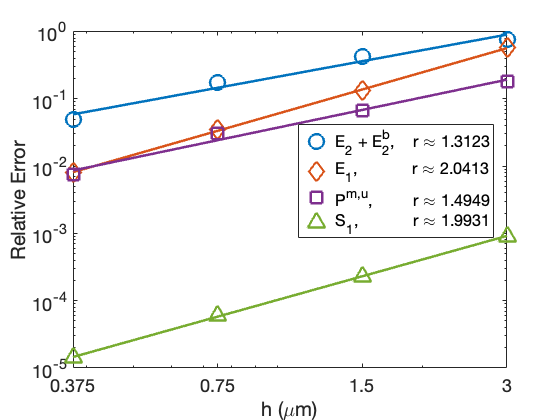}
}
\caption{Mesh convergence study in a rectangular domain. Errors are calculated at $t = 5$ seconds with respect to a fine mesh solution $\psi_\text{fine}$ ($h = 0.1875 \; \mu$m), which serves as a reference solution. The approximate convergence rate $r$ is displayed in the legend for each species.} \label{fig:meshConvergence}
\end{figure}

In Figure \ref{fig:meshConvergence}, we present the errors at time $t = 5$ seconds for four different species that represent various components of the clotting model:
\begin{enumerate}
    \item $S_1$ is a fluid-phase biochemical species that enters the domain with a fixed concentration and reacts on subendothelium with $E_0$ to produce the enzyme $E_1$.
    \item $E_1$ is a fluid-phase species that enters the domain through a reactive boundary condition on the subendothelium. 
    \item $P^{m,u}$ is fluid-phase platelet species that enters the domain with a marginated profile and experiences hindered transport in the presence of a growing thrombus.  This species has reaction terms that depend on the adhesion region and biochemical agonists such as ADP and thrombin. 
    \item $E_2+E_2^b$ represents the total thrombin concentration and serves as a surrogate measure of overall clot formation. $E_2$ is a fluid-phase species, while $E_2^b$ is platelet-bound.
\end{enumerate}
The error plot in Figure \ref{fig:meshConvergence} demonstrates that the fluid-phase biochemical species converge with second-order accuracy, while the platelet-related species exhibit a super-linear convergence rate.

\section*{Effect of mesh topology}
\begin{figure}[H]
\makebox[\linewidth][c]{%
\begin{subfigure}[b]{.50\textwidth}
  \centering
  \includegraphics[width=0.95\textwidth,trim={5cm 8.5cm 5cm 8.5cm},clip]{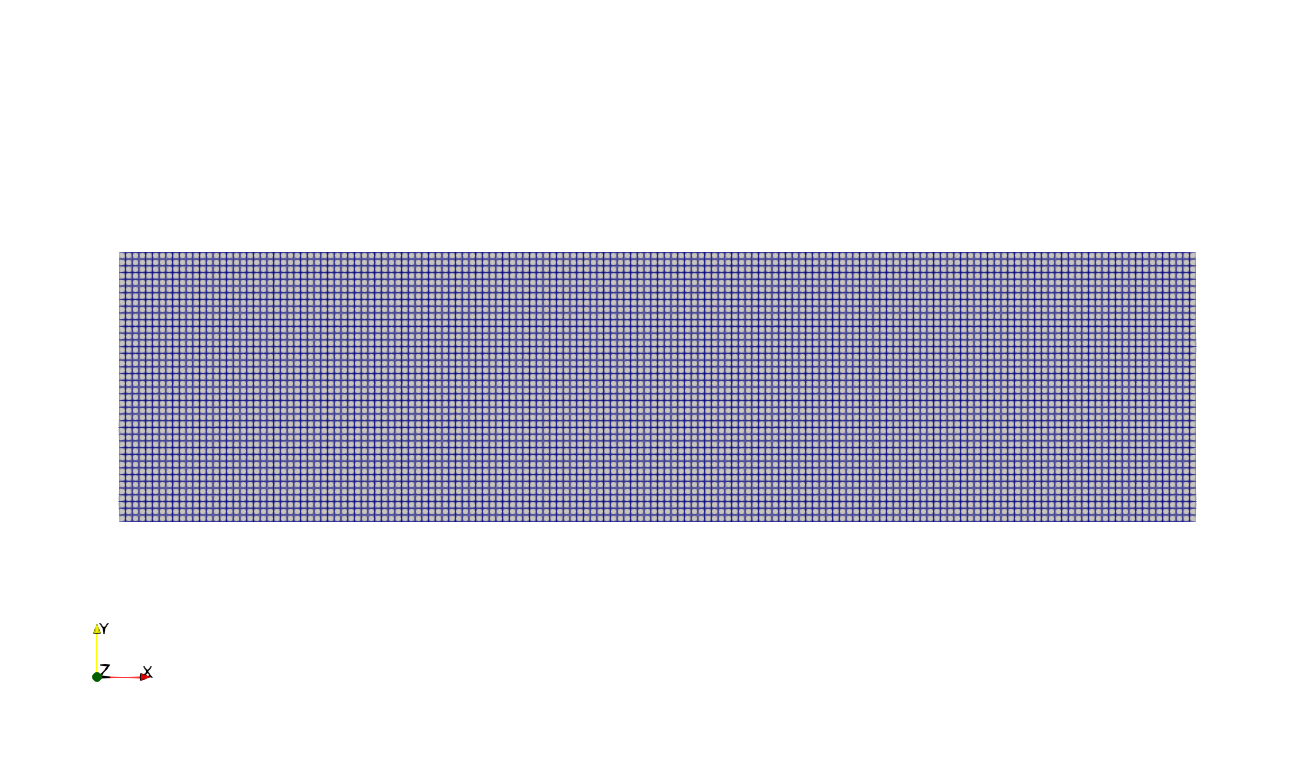}
\end{subfigure}%

\begin{subfigure}[b]{.50\textwidth}
  \centering
  \includegraphics[width=0.95\textwidth,trim={5cm 8.5cm 5cm 8.5cm},clip]{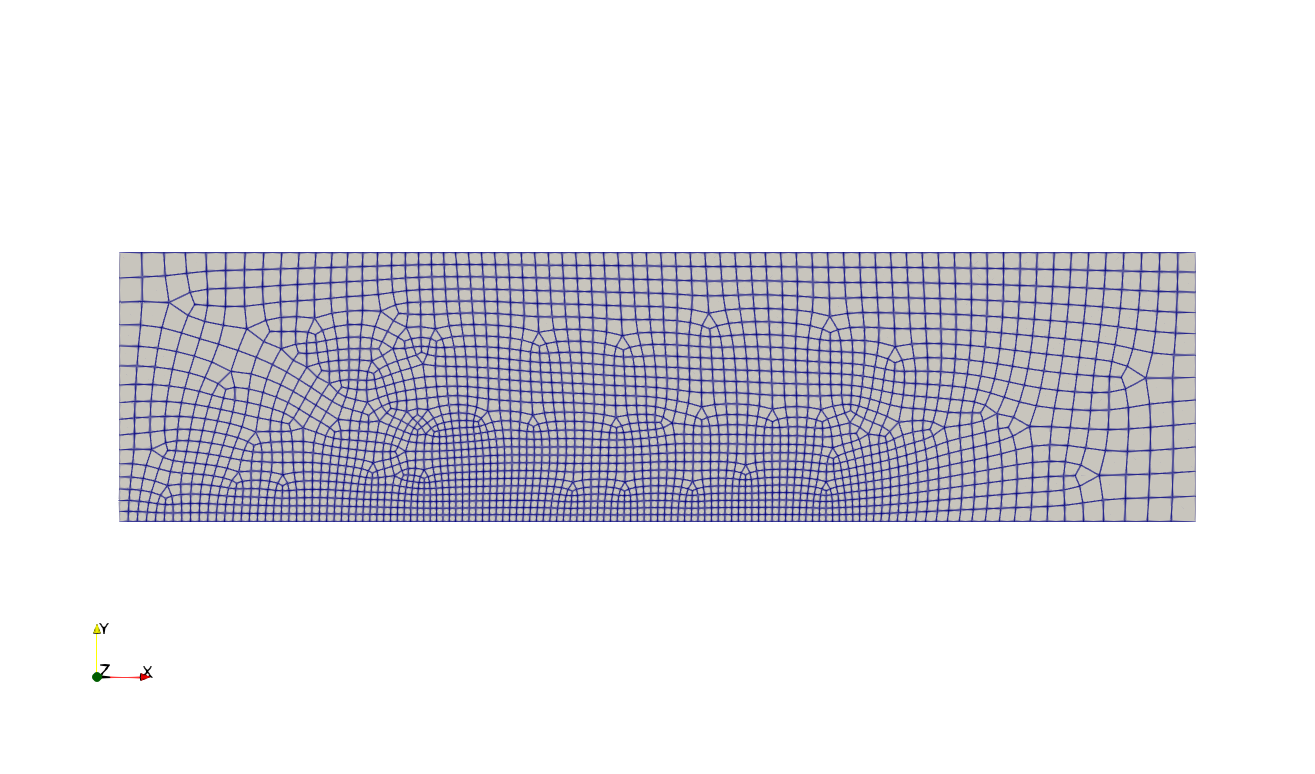}
\end{subfigure}%
}

\makebox[\linewidth][c]{%
\begin{subfigure}[b]{.50\textwidth}
  \centering
  \includegraphics[width=0.98\textwidth,trim={1cm 8cm 3.4cm 8.5cm},clip]{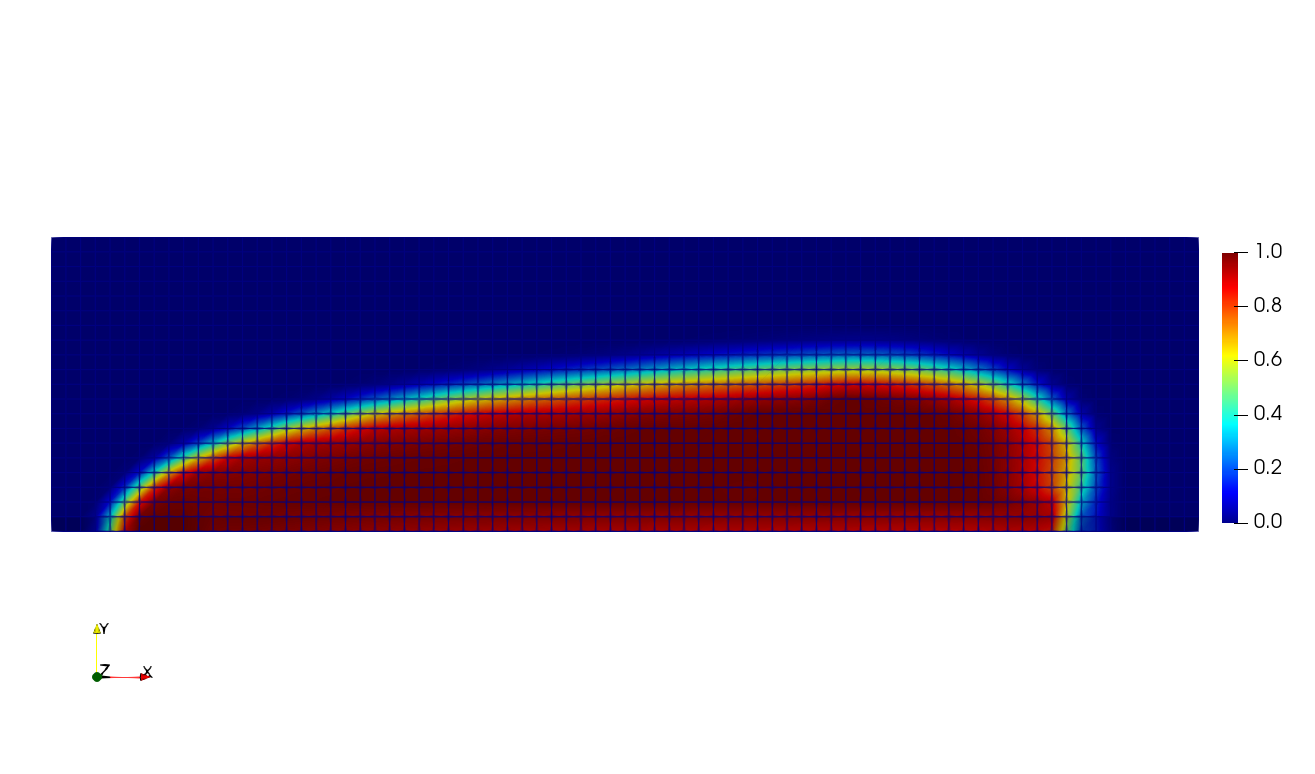}
\end{subfigure}%

\begin{subfigure}[b]{.50\textwidth}
  \centering
  \includegraphics[width=0.98\textwidth,trim={1cm 8cm 3.4cm 8.5cm},clip]{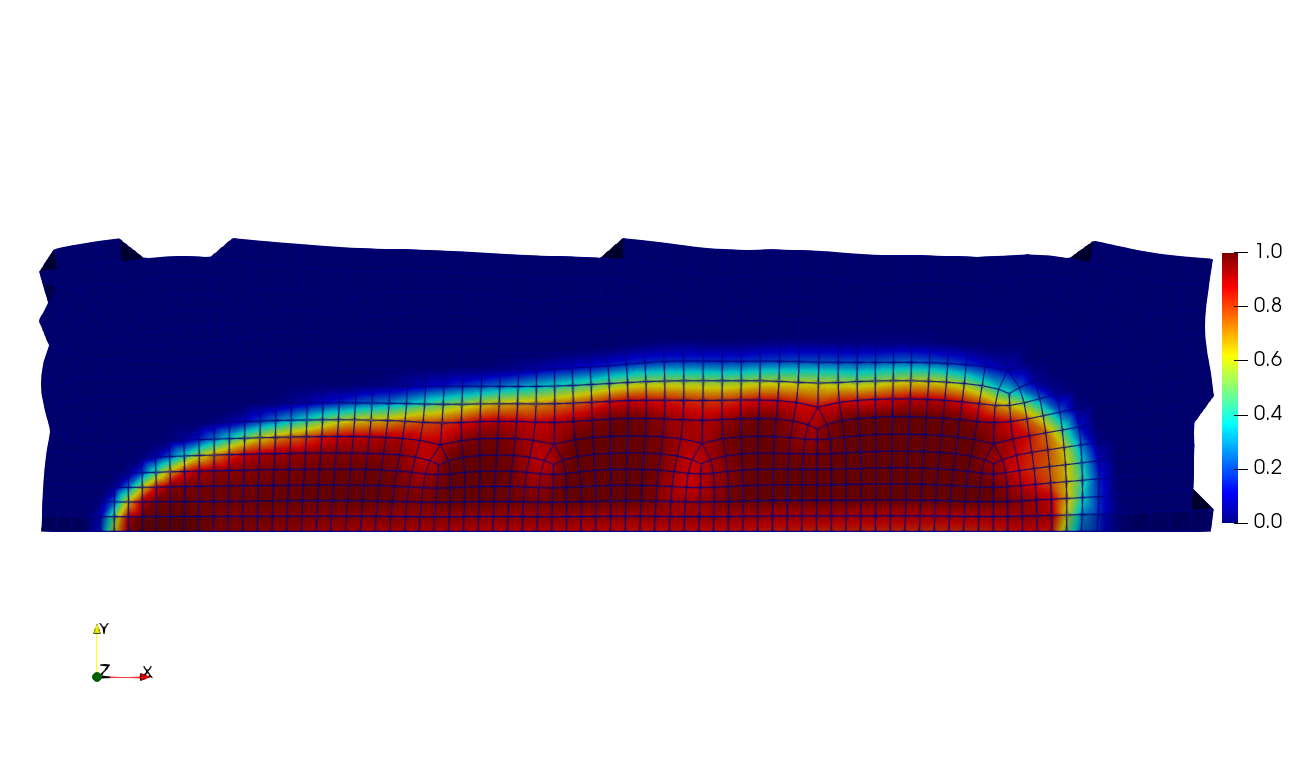}
\end{subfigure}%
}

\makebox[\linewidth][c]{%
\begin{subfigure}[b]{.50\textwidth}
  \centering
  \includegraphics[width=0.98\textwidth,trim={0cm 0cm 0cm 0cm},clip]{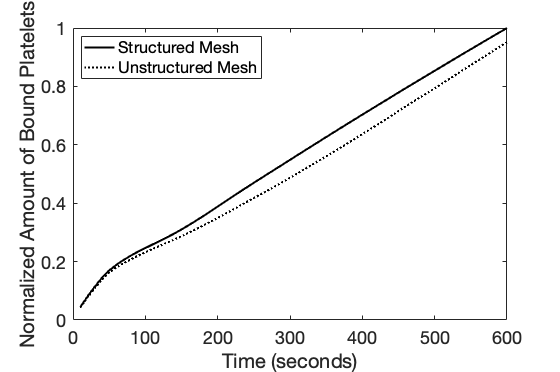}
\end{subfigure}%

\begin{subfigure}[b]{.50\textwidth}
  \centering
  \includegraphics[width=\textwidth,trim={0cm 0cm 0cm 0cm},clip]{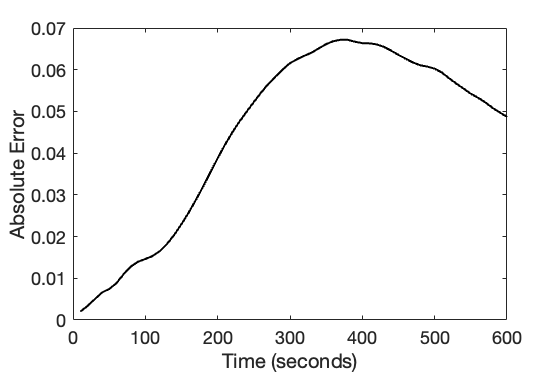}
\end{subfigure}%
}
\caption{Comparison of the effect of mesh topology on thrombus growth for a rectangular domain measuring 240 $\mu$m by 60 $\mu$m. The top row depicts a structured (uniform) mesh on the left and an unstructured mesh on the right. The second row provides a close-up view of the bound platelet fraction in a 120 $\mu$m long by 30 $\mu$m high region surrounding the thrombus after 600 s of simulation. The bottom row quantifies the discrepancy between the simulations by integrating the bound platelet fraction over the entire domain and normalizing it by the maximum value of that quantity.} \label{fig:Gmsh}
\end{figure}

To investigate the impact of mesh topology on thrombus formation, we generated an unstructured mesh using the open-source software Gmsh, as shown in Figure \ref{fig:Gmsh}. Our preliminary findings suggest that the specific mesh topology has a limited effect on the overall growth of the thrombus. However, we observed variations in clot density, with less dense regions occurring in areas where the mesh topology is more skewed. It is important to note that these observations are based on a single test with an unconventional mesh, and further investigations incorporating carefully selected discretization schemes may help mitigate this issue.

To assess the influence of mesh topology on thrombus growth, we conducted a comparative study using two different mesh configurations within a rectangular domain, as depicted in Figure \ref{fig:Gmsh}. The left column represents simulations conducted using a structured mesh, while the right column corresponds to simulations conducted using an unstructured mesh. Our comparison revealed similar patterns of thrombus growth in both simulations. We quantitatively evaluated the differences in thrombi by calculating the integral of the bound platelet fraction over the domain and normalizing it by its maximum value, resulting in the "normalized amount of bound platelets." The bottom plots in Figure \ref{fig:Gmsh} demonstrate that the error in the normalized amount of bound platelets is less than 7\% between the structured and unstructured mesh configurations after 600 seconds of simulation.  

\section*{Quantitative validation}
To quantitatively validate the \emph{clotFoam} solver, we conducted comparative studies with the works of Leiderman and Fogelson \cite{Leiderman2011grow} and Schoeman et al. \cite{Schoeman2016}. It is important to acknowledge that the clotting model implemented in \emph{clotFoam} is a significantly reduced version of the Leiderman and Fogelson model. This reduction was intended to provide researchers with a flexible framework that can be adapted to suit their specific research needs and objectives. As a result, direct comparisons between these models can be challenging due to inherent differences in their formulations.  However, considering the results obtained from these comparative studies, along with the mesh convergence analysis depicted in Figure \ref{fig:meshConvergence}, it becomes evident that \emph{clotFoam} demonstrates reliability in simulating thrombus growth under arterial flow conditions. 

\subsubsection*{Comparison with Leiderman and Fogelson 2011}
\begin{figure}[H]
\makebox[\linewidth][c]{%
\begin{subfigure}[b]{.33\textwidth}
  \centering
  \includegraphics[width=\textwidth,trim={0cm 0cm 0cm 0cm},clip]{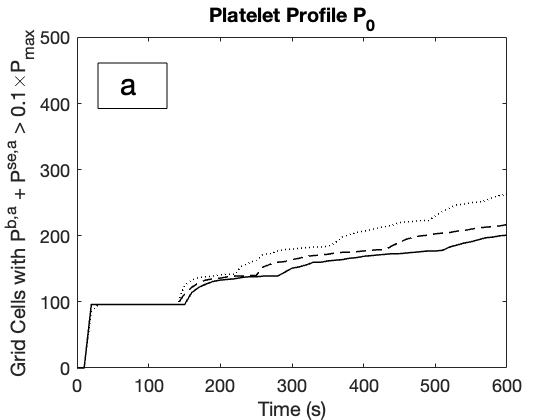}
\end{subfigure}%

\begin{subfigure}[b]{.33\textwidth}
  \centering
  \includegraphics[width=\textwidth,trim={0cm 0cm 0cm 0cm},clip]{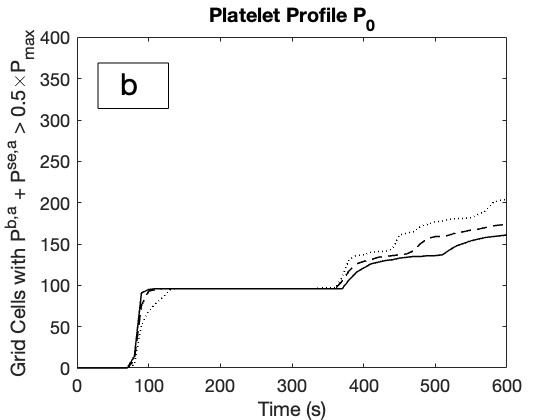}
\end{subfigure}%

\begin{subfigure}[b]{.33\textwidth}
  \centering
  \includegraphics[width=\textwidth,trim={0cm 0cm 0cm 0cm},clip]{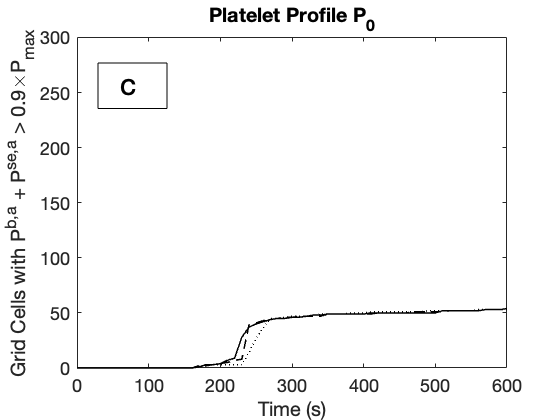}
\end{subfigure}%
}

\makebox[\linewidth][c]{%
\begin{subfigure}[b]{.33\textwidth}
  \centering
  \includegraphics[width=\textwidth,trim={0cm 0cm 0cm 0cm},clip]{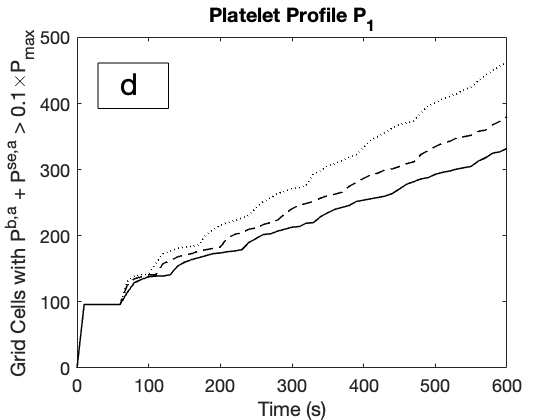}
\end{subfigure}%

\begin{subfigure}[b]{.33\textwidth}
  \centering
  \includegraphics[width=\textwidth,trim={0cm 0cm 0cm 0cm},clip]{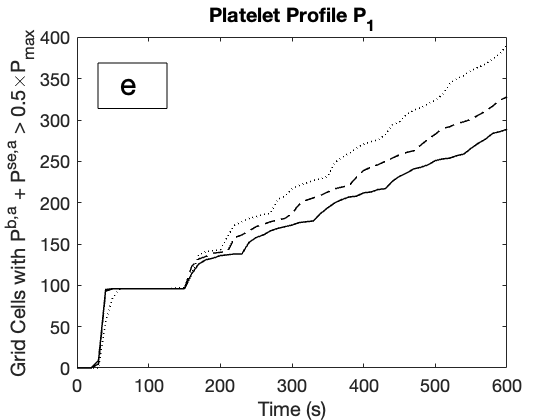}
\end{subfigure}%

\begin{subfigure}[b]{.33\textwidth}
  \centering
  \includegraphics[width=\textwidth,trim={0cm 0cm 0cm 0cm},clip]{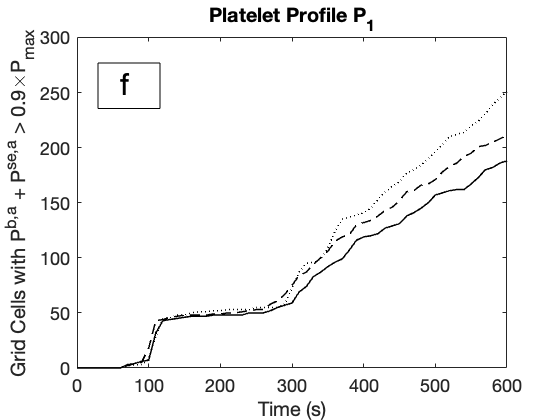}
\end{subfigure}%
}

\makebox[\linewidth][c]{%
\begin{subfigure}[b]{.33\textwidth}
  \centering
  \includegraphics[width=\textwidth,trim={0cm 0cm 0cm 0cm},clip]{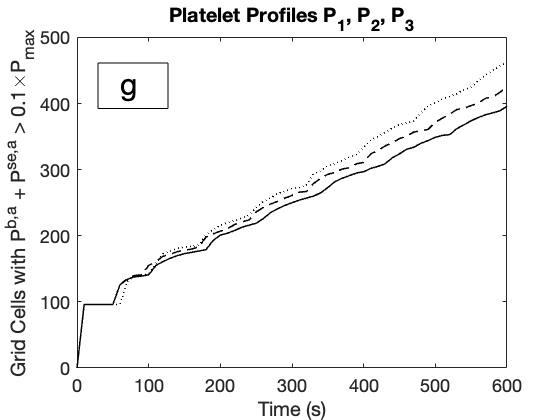}
\end{subfigure}%

\begin{subfigure}[b]{.33\textwidth}
  \centering
  \includegraphics[width=\textwidth,trim={0cm 0cm 0cm 0cm},clip]{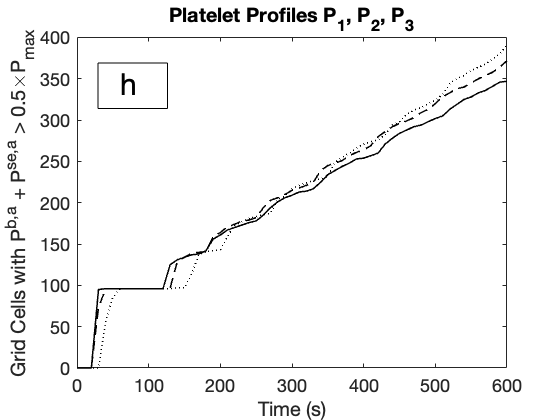}
\end{subfigure}%

\begin{subfigure}[b]{.33\textwidth}
  \centering
  \includegraphics[width=\textwidth,trim={0cm 0cm 0cm 0cm},clip]{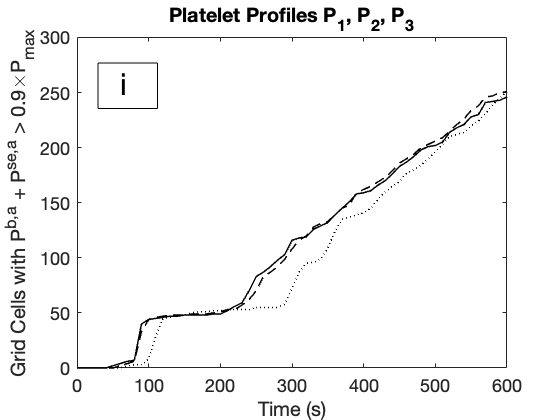}
\end{subfigure}%
}

\caption{Quantitative analysis of clot 'area' for comparison with Fig. 8 in Leiderman \& Fogelson \cite{Leiderman2011grow}. The top row (a-c) presents results obtained using a uniform platelet profile. The middle row (d-f) shows results obtained using a fixed nonuniform platelet profile $P_1$, as defined in Fig. 3 of Leiderman \& Fogelson \cite{Leiderman2011grow}. The bottom row (g–i) displays results obtained using platelet profiles $P_1$, $P_2$, and $P_3$ for wall shear rates of 500, 1000, and 1500 s$^{-1}$, respectively. In all cases, the dotted line, dashed line, and solid line represent wall shear rates of 500, 1000, and 1500 s$^{-1}$, respectively.} \label{fig:thrombusCellCounts}
\end{figure}

In the first comparative study, we compared thrombus growth from the \emph{clotFoam} solver to Figure 8 in Leiderman and Fogelson \cite{Leiderman2011grow}. We simulated thrombus growth under three shear rates (500, 1000, and 1500 s$^{-1}$) with four different inlet platelet profiles ($P_0$, $P_1$, $P_2$, $P_3$ as defined in Figure 3 of Leiderman and Fogelson) that have increasing margination ratios. To quantify the differences, we computed the 'area' in which the bound platelet concentration exceeded 10\%, 50\%, and 90\% of $P_\text{max}$, respectively. The first percentage provides a measure of the overall thrombus size, while the latter two percentages provide information about the bound platelet density distribution within the thrombus. These 'areas' were computed in ParaView by counting the grid cells in which the various platelet concentration levels were exceeded, and are presented in Figure \ref{fig:thrombusCellCounts}.

Similar to the plots in Leiderman and Fogelson \cite{Leiderman2011grow}, we observed the appearance of a temporary plateau at an area of 100 grid cells. This corresponds to the region where the adhesion rate function $k_\text{adh}H_\text{adh}(x)$ is non-zero. However, for the highest concentration of bound platelets, when the bound platelets concentration exceeded 90\% of $P_\text{max}$, we observed a plateau at an area of 50 grid cells. This indicates that only half of the adhesion region exceeds a density of 90\% of $P_\text{max}$. This behavior can be seen in Figure \ref{fig:exampleThrombosis} of our manuscript, where the adhesion region at the bottom of the thrombus has a less dense concentration compared to the rest of the clot. Given the reduced nature of the \emph{clotFoam} model, it is not surprising that our results for the timing of plateaus, overall clot growth, and shear rate dynamics differ from those published by Leiderman and Fogelson. However, we do observe similar characteristics of thrombus growth, with the curves in Figure \ref{fig:thrombusCellCounts} following the same general trajectories as those reported by Leiderman and Fogelson.

\subsubsection*{Comparsion with Schoeman et al. 2016}
In the second comparative study, we conducted an assessment of thrombus formation 
over time by comparing our results with Figure 5 of Schoeman et al. \cite{Schoeman2016}. Our primary objective with the illustrative example of hemostasis in a microfluidic device using the \emph{clotFoam} software was to demonstrate its capability in simulating clot formation within such a device. However, it is important to acknowledge that the platelet model employed by \emph{clotFoam} does not incorporate crucial factors like von Willebrand factor (vWF) or shear dependence in the platelet aggregation process, which are known to play significant roles in high shear settings, such as those encountered in the H-shaped microfluidic device described by Schoeman et al.

\begin{figure}[H]
\makebox[\linewidth][c]{%
  \includegraphics[width=0.5\textwidth,]{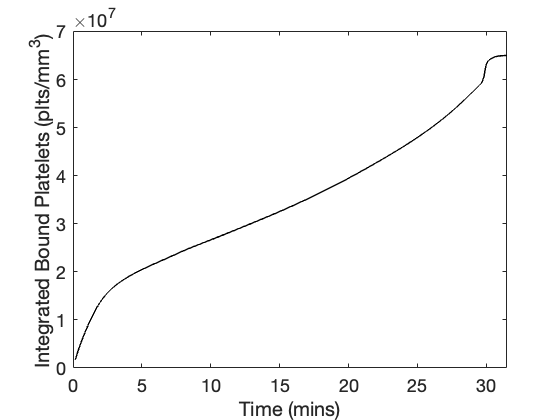}
}
\caption{Characteristic platelet accumulation in the injury channel for whole blood on collagen-TF.  The plot shows the progressive accumulation of platelets over time until the injury channel reaches occlusion at approximately 30 minutes.} \label{fig:H-domain-analysis}
\end{figure}

Despite the limitations of the platelet aggregation model in \emph{clotFoam}, Figure \ref{fig:H-domain-analysis} illustrates characteristic platelet accumulation in the injury channel for whole blood on collagen and tissue factor (TF). The plot shows that as time progresses, the platelet accumulation increases until the injury channel reaches full occlusion at approximately 30 minutes, resulting in a plateau in the amount of bound platelets. This behavior is similar to what is observed in the Schoeman et al. paper, although the time scales differ. These findings demonstrate that the base clotting model in \emph{clotFoam} is capable of simulating hemostasis in a microfluidic device. Researchers can easily modify the model to incorporate the effects of shear, von Willebrand factor (vWF), thromboxane A2, fibrin, and other relevant biochemical species involved in the coagulation process. The primary objective of developing \emph{clotFoam} was to provide researchers with a versatile framework that can be easily customized and adapted to different clotting models. This flexibility allows for further exploration and customization to meet the specific research requirements of various studies in the field of hemostasis and thrombosis.

\bibliographystyle{elsarticle-num} 
\bibliography{Bibliography}

\begin{thebibliography}{10}
\expandafter\ifx\csname url\endcsname\relax
  \def\url#1{\texttt{#1}}\fi
\expandafter\ifx\csname urlprefix\endcsname\relax\def\urlprefix{URL }\fi
\expandafter\ifx\csname href\endcsname\relax
  \def\href#1#2{#2} \def\path#1{#1}\fi

\bibitem{FOGELSON1998}
A.~L. Fogelson, A.~L. Kuharsky, Membrane binding-site density can modulate
  activation thresholds in enzyme systems, Journal of Theoretical Biology
  193~(1) (1998) 1--18.
\newblock \href {https://doi.org/10.1006/jtbi.1998.0670}
  {\path{doi:10.1006/jtbi.1998.0670}}.

\bibitem{kuharsky_surface-mediated_2001}
A.~L. Kuharsky, A.~L. Fogelson, Surface-{Mediated} {Control} of {Blood}
  {Coagulation}: {The} {Role} of {Binding} {Site} {Densities} and {Platelet}
  {Deposition}, Biophysical Journal 80~(3) (2001) 1050--1074.
\newblock \href {https://doi.org/10.1016/S0006-3495(01)76085-7}
  {\path{doi:10.1016/S0006-3495(01)76085-7}}.

\bibitem{Leiderman2011grow}
K.~Leiderman, A.~L. Fogelson, Grow with the flow: a spatial--temporal model of
  platelet deposition and blood coagulation under flow, Mathematical medicine
  and biology: a journal of the IMA 28~(1) (2011) 47--84.
\newblock \href {https://doi.org/10.1093/imammb/dqq005}
  {\path{doi:10.1093/imammb/dqq005}}.

\bibitem{miyazawa2023inhibitionB}
K.~Miyazawa, A.~L. Fogelson, K.~Leiderman, Inhibition of platelet-surface-bound
  proteins during coagulation under flow i: Antithrombin and heparin,
  Biophysical journal 122~(1) (2023) 230--240.
\newblock \href {https://doi.org/10.1016/j.bpj.2022.11.023}
  {\path{doi:10.1016/j.bpj.2022.11.023}}.

\bibitem{NeevesKeithB2016MMoH}
K.~B. Neeves, K.~Leiderman, Mathematical models of hemostasis, Trauma induced
  coagulopathy (2016) 567--584\href
  {https://doi.org/10.1007/978-3-319-28308-1\_35}
  {\path{doi:10.1007/978-3-319-28308-1\_35}}.

\bibitem{Leidermann2018Chp}
K.~Leiderman, B.~Bannish, M.~Kelley, A.~Palmisano, Mathematical models of
  thrombus formation and fibrinolysis, in: Cardiovascular thrombus: from
  pathology and clinical presentation to imaging, pharmacotherapy and
  interventions, Academic Press San Diego, 2018, pp. 67--82.
\newblock \href {https://doi.org/10.1016/B978-0-12-812615-8.00005-3}
  {\path{doi:10.1016/B978-0-12-812615-8.00005-3}}.

\bibitem{Diamond2013Review}
S.~L. Diamond, Systems biology of coagulation, Journal of Thrombosis and
  Haemostasis 11 (2013) 224--232.
\newblock \href {https://doi.org/10.1111/jth.12220}
  {\path{doi:10.1111/jth.12220}}.

\bibitem{YesudasanSumith2019Raic}
S.~Yesudasan, R.~D. Averett, Recent advances in computational modeling of
  fibrin clot formation: A review, Computational biology and chemistry 83
  (2019) 107148.
\newblock \href {https://doi.org/10.1016/j.compbiolchem.2019.107148}
  {\path{doi:10.1016/j.compbiolchem.2019.107148}}.

\bibitem{AnandM.2022Cmoh}
M.~Anand, M.~Panteleev, F.~Ataullakhanov, Computational models of hemostasis:
  Degrees of complexity, Applications in Engineering Science 10 (2022) 100103.
\newblock \href {https://doi.org/10.1016/j.apples.2022.100103}
  {\path{doi:10.1016/j.apples.2022.100103}}.

\bibitem{TaylorJoshuaO.2016Doac}
J.~O. Taylor, R.~S. Meyer, S.~Deutsch, K.~B. Manning, Development of a
  computational model for macroscopic predictions of device-induced thrombosis,
  Biomechanics and modeling in mechanobiology 15 (2016) 1713--1731.
\newblock \href {https://doi.org/10.1007/s10237-016-0793-2}
  {\path{doi:10.1007/s10237-016-0793-2}}.

\bibitem{Govindarajan2018}
V.~Govindarajan, S.~Zhu, R.~Li, Y.~Lu, S.~L. Diamond, J.~Reifman, A.~Y.
  Mitrophanov, Impact of tissue factor localization on blood clot structure and
  resistance under venous shear, Biophysical journal 114~(4) (2018) 978--991.
\newblock \href {https://doi.org/10.1016/j.bpj.2017.12.034}
  {\path{doi:10.1016/j.bpj.2017.12.034}}.

\bibitem{Rojano2022}
R.~M{\'e}ndez~Rojano, M.~Zhussupbekov, J.~F. Antaki, D.~Lucor, Uncertainty
  quantification of a thrombosis model considering the clotting assay
  pfa-100{\textregistered}, International Journal for Numerical Methods in
  Biomedical Engineering 38~(5) (2022) e3595.
\newblock \href {https://doi.org/doi.org/10.1002/cnm.3595}
  {\path{doi:doi.org/10.1002/cnm.3595}}.

\bibitem{BouchnitaAnass2021Mcmo}
A.~Bouchnita, A.~V. Belyaev, V.~Volpert, Multiphase continuum modeling of
  thrombosis in aneurysms and recirculation zones, Physics of Fluids 33~(9)
  (2021) 093314.
\newblock \href {https://doi.org/10.1063/5.0057393}
  {\path{doi:10.1063/5.0057393}}.

\bibitem{Schoeman2016}
R.~M. Schoeman, K.~Rana, N.~Danes, M.~Lehmann, J.~A. Di~Paola, A.~L. Fogelson,
  K.~Leiderman, K.~B. Neeves, A microfluidic model of hemostasis sensitive to
  platelet function and coagulation, Cellular and molecular bioengineering 10
  (2017) 3--15.
\newblock \href {https://doi.org/10.1007/s12195-016-0469-0}
  {\path{doi:10.1007/s12195-016-0469-0}}.

\bibitem{Danes}
N.~A. Danes, K.~Leiderman, A density-dependent fem-fct algorithm with
  application to modeling platelet aggregation, International journal for
  numerical methods in biomedical engineering 35~(9) (2019) e3212.
\newblock \href {https://doi.org/10.1002/cnm.3212}
  {\path{doi:10.1002/cnm.3212}}.

\bibitem{Leiderman2012}
K.~Leiderman, A.~L. Fogelson, The influence of hindered transport on the
  development of platelet thrombi under flow, Bulletin of mathematical biology
  75 (2013) 1255--1283.
\newblock \href {https://doi.org/10.1007/s11538-012-9784-3}
  {\path{doi:10.1007/s11538-012-9784-3}}.

\bibitem{Rezaeimoghaddam2022Cmot}
M.~Rezaeimoghaddam, F.~N. van~de Vosse, Continuum modeling of thrombus
  formation and growth under different shear rates, Journal of Biomechanics 132
  (2022) 110915.
\newblock \href {https://doi.org/10.1016/j.jbiomech.2021.110915}
  {\path{doi:10.1016/j.jbiomech.2021.110915}}.

\bibitem{Wu2017}
W.-T. Wu, M.~A. Jamiolkowski, W.~R. Wagner, N.~Aubry, M.~Massoudi, J.~F.
  Antaki, Multi-constituent simulation of thrombus deposition, Scientific
  reports 7~(1) (2017) 1--16.
\newblock \href {https://doi.org/10.1038/srep42720}
  {\path{doi:10.1038/srep42720}}.

\bibitem{OpenFOAM-user-ref}
C.~Greenshields, \href{https://doc.cfd.direct/openfoam/user-guide-v9}{OpenFOAM
  v9 User Guide}, The OpenFOAM Foundation, London, UK, 2021.

\bibitem{shankar2022three}
K.~N. Shankar, Y.~Zhang, T.~Sinno, S.~L. Diamond, A three-dimensional
  multiscale model for the prediction of thrombus growth under flow with
  single-platelet resolution, PLOS Computational Biology 18~(1) (2022)
  e1009850.
\newblock \href {https://doi.org/10.1371/journal.pcbi.1009850}
  {\path{doi:10.1371/journal.pcbi.1009850}}.

\bibitem{mendez2022fibrin}
R.~M{\'e}ndez~Rojano, A.~Lai, M.~Zhussupbekov, G.~W. Burgreen, K.~Cook, J.~F.
  Antaki, A fibrin enhanced thrombosis model for medical devices operating at
  low shear regimes or large surface areas, PLOS Computational Biology 18~(10)
  (2022) e1010277.
\newblock \href {https://doi.org/10.1371/journal.pcbi.1010277}
  {\path{doi:10.1371/journal.pcbi.1010277}}.

\bibitem{ISSA1986-PISO}
R.~I. Issa, Solution of the implicitly discretised fluid flow equations by
  operator-splitting, Journal of computational physics 62~(1) (1986) 40--65.
\newblock \href {https://doi.org/10.1016/0021-9991(86)90099-9}
  {\path{doi:10.1016/0021-9991(86)90099-9}}.

\bibitem{feletou2011endothelium}
M.~F{\'{e}}l{\'{e}}tou, The endothelium, part i: Multiple functions of the
  endothelial cells -- focus on endothelium-derived vasoactive mediators,
  Colloquium Series on Integrated Systems Physiology: From Molecule to Function
  3~(4) (2011) 1--306.
\newblock \href {https://doi.org/10.4199/c00031ed1v01y201105isp019}
  {\path{doi:10.4199/c00031ed1v01y201105isp019}}.

\bibitem{eckstein1991model}
E.~C. Eckstein, F.~Belgacem, Model of platelet transport in flowing blood with
  drift and diffusion terms, Biophysical journal 60~(1) (1991) 53--69.
\newblock \href {https://doi.org/10.1016/S0006-3495(91)82030-6}
  {\path{doi:10.1016/S0006-3495(91)82030-6}}.

\bibitem{VANLEER1974361}
B.~{van Leer}, Towards the ultimate conservative difference scheme. ii.
  monotonicity and conservation combined in a second-order scheme, Journal of
  Computational Physics 14~(4) (1974) 361--370.
\newblock \href {https://doi.org/10.1016/0021-9991(74)90019-9}
  {\path{doi:10.1016/0021-9991(74)90019-9}}.

\bibitem{OpenFOAM-wiki-grading}
Scripts/blockmesh grading calculation,
  \url{https://openfoamwiki.net/index.php/Scripts/blockMesh\_grading\_calculation}
  (2020).

\end{thebibliography}

\end{document}